\documentclass[%
 reprint,
 amsmath,amssymb,
 aps,
floatfix,
]{revtex4-2}
\pdfoutput=1

\usepackage{graphicx}
\usepackage{dcolumn}
\usepackage{bm}

\usepackage[colorlinks=true,linkcolor=blue,citecolor=blue,allcolors=blue]{hyperref}

\begin{document}

\preprint{APS/123-QED}
\title{Capillary and Viscous Fracturing During Drainage in Porous Media}
\author{Francisco J. Carrillo}
\homepage{https://github.com/Franjcf}
\affiliation{%
 Department of Chemical and Biological Engineering, Princeton University, Princeton, NJ, USA
}%

\author{Ian C. Bourg}
\homepage{http://bourg.princeton.edu}
\affiliation{
 Department of Civil and Environmental Engineering, Princeton University, Princeton, NJ, USA \\
}%
\affiliation{
High Meadows Environmental Institute, Princeton University, Princeton, NJ, USA
}%

\date{\today}%

\begin{abstract}
Detailed understanding of the couplings between fluid flow and solid deformation in porous media is crucial for the development of novel technologies relating to a wide range of geological and biological processes. A particularly challenging phenomenon that emerges from these couplings is the transition from fluid invasion to fracturing during multiphase flow. Previous studies have shown that this transition is highly sensitive to fluid flow rate, capillarity, and the structural properties of the porous medium. However, a comprehensive characterization of the relevant fluid flow and material failure regimes does not exist. Here, we used our newly developed Multiphase Darcy-Brinkman-Biot framework to examine the transition from drainage to material failure during viscously-stable multiphase flow in soft porous media in a broad range of flow, wettability, and solid rheology conditions. We demonstrate the existence of three distinct material failure regimes controlled by non-dimensional numbers that quantify the balance of viscous, capillary, and structural forces in the porous medium.

\end{abstract}

\maketitle

\section{Introduction}

Multiphase flow in deformable porous media is a ubiquitous phenomenon in natural and engineered systems that underlies key processes in water and energy resource engineering and materials science, including membrane filtration, soil wetting/drying, unconventional hydrocarbon recovery, and geologic carbon sequestration \cite{Bacher2019,Rass2018,Towner1987}. A key obstacle to more accurate representations of this phenomenon is our limited understanding of the transition from fluid invasion to flow-induced fracturing, i.e., material failure caused by multiphase flow. In large part, this limitation is caused by a lack of computational approaches capable of representing multiphase flow in fractured deformable porous media. 

Previous work on \textit{multiphase} flow within \textit{static} porous media is extensive and includes detailed examinations of the influence of wettability, viscosity, and flow rate on flow in unsaturated porous media at multiple scales. In particular, existing studies have demonstrated how capillary forces give rise to differences between drainage and imbibition \cite{Lenormand1986}; how the ratio of fluid viscosities controls the stability of the invading fluid front \cite{JorgenMalby1985,Saffman_1958,Stokes1976}; and how the magnitude of the capillary number delineates distinct flow regimes \cite{Ferer2004,Yortsos1997}. Each of the aforementioned controls is highly dependent on the system of interest. This complicates efforts to develop general relative permeability and capillary pressure models that apply to most systems of interest \cite{Picchi2018,Picchi2019,RHBrooks1964,VanGenutchen1980}.

Flow of a \textit{single fluid phase} through \textit{deformable} porous media also has been studied in depth. Numerical modeling studies are largely based on the work of Biot and Terzaghi \cite{Biot1941,Terzaghi1943} and have been used to reproduce the behavior of arteries, boreholes, swelling clays, and gels \cite{Auton2017a,Bertrand2016b,Carrillo2019a,MacMinn2015a}. In the last decade, fundamental studies have generated detailed information on the dynamics that arise from fluid-solid couplings beyond the ideal poroelastic regime, including fracturing, granular fingering, and frictional fingering \cite{Campbell2017,Sandnes2011a,Zhang2013}. In particular, these studies have shown that the main parameters controlling the deformation of a porous solid by single phase flow are the material softness and the magnitude of the fluid-solid momentum transfer \cite{Sandnes2011a}.

The study of \textit{multiphase} flow in a \textit{deformable} porous medium is inherently more complex than the problems outlined above, as it requires simultaneous consideration of capillarity, wetting dynamics, fluid rheology, and solid deformation. Deformation modes associated with material failure (i.e., multiphase fracturing) are particularly challenging as they require simultaneous representation of multiphase flow in fractures and in the surrounding porous matrix. The existing detailed examinations of this phenomenon have focused exclusively on granular systems. Notably, Holtzman \& Juanes \cite{Holtzman2010a,Holtzman2012a} used experiments and discrete element models to demonstrate that the transitions between capillary fingering,
viscous fingering, and fracturing during multiphase flow in granular media reflect two non-dimensional numbers: a fracturing number (ratio of fluid
driving force to solid cohesive force) and a modified capillary number (the ratio between viscous and capillary pressure drops). Other discrete element approaches have shown that fracturing is highly dependent on the invading fluid's capillary entry pressure
\cite{Jain2009,Meng2020}. However, it is not clear how these conclusions translate to continuous non-granular systems.

To the best of our knowledge, no
experimental or numerical investigation has simultaneously explored the effects of flow rate, wettability, and
deformability during multiphase flow in deformable porous media at the \textit{continuum scale} and identified the controlling
parameters that relate \textit{all three} properties within a single phase diagram. Here, we use simulations carried out with our new Multiphase Darcy-Brinkman-Biot (DBB) framework \cite{Carrillo2020MDBB} to fill this knowledge gap and identify non-dimensional parameters that govern viscously-stable fluid drainage and fracturing in deformable porous media. We also find that the fracturing dynamics predicted by our continuum-scale framework is consistent with those observed or predicted for granular systems. In other words, in systems with a large length scale separation between pores and fractures, volume-averaged properties are sufficient to capture the onset and propagation of fractures at the continuum scale.

\section{Modeling Framework}

\begin{figure} 
\includegraphics[width=0.483\textwidth]{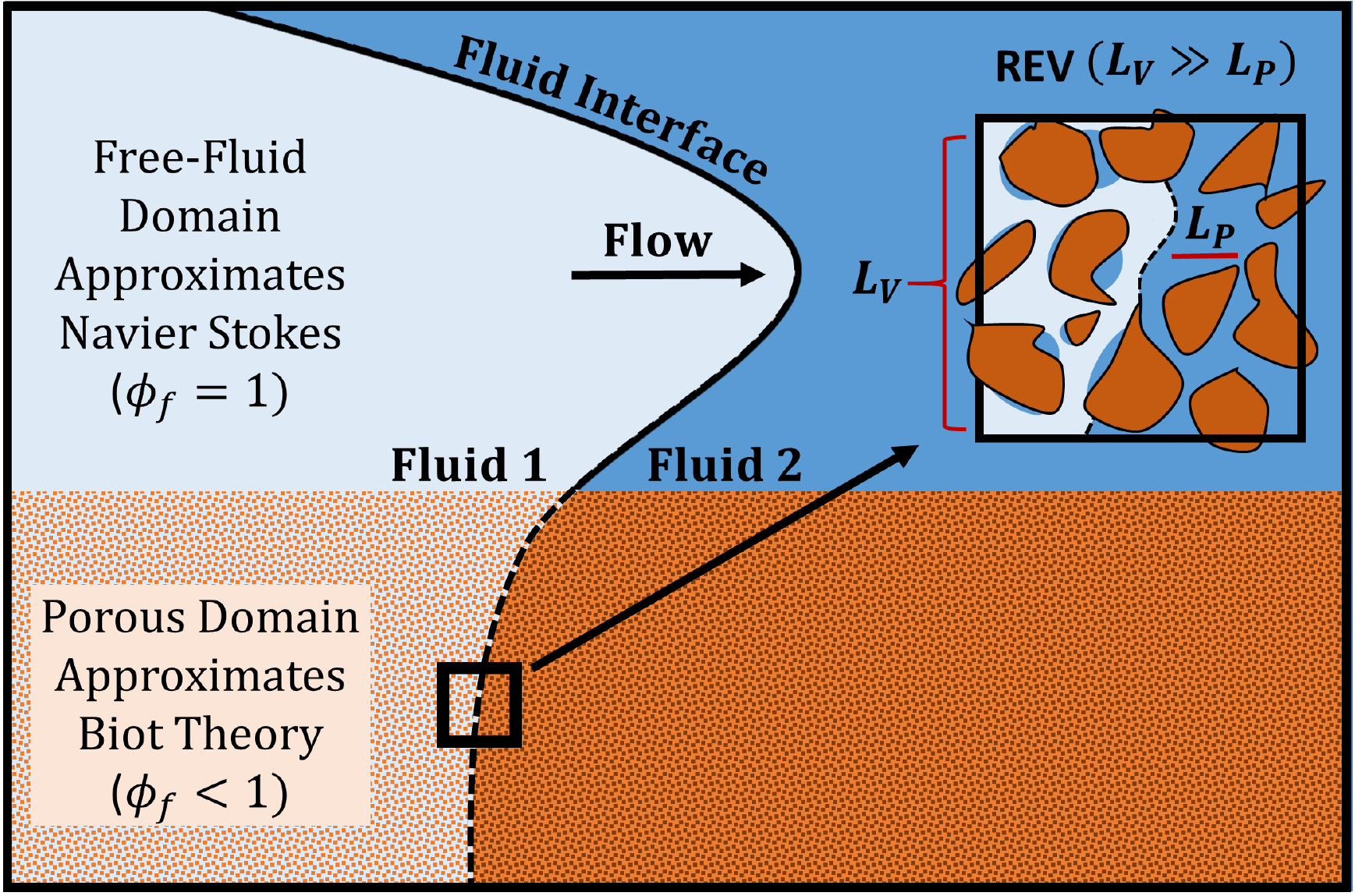}
\caption{\label{fig:conceptual} Conceptual representation of the Multiphase DBB framework. The porous domain is shown in the lower half, the free-fluid domain is shown in the upper half, the two immiscible fluids (left and right) are shown in different shades of blue and are separated by an interface (black), and ${\phi }_f$ is the porosity. REV is the ``Representative Elementary Volume'' over which all equations are averaged.}
\end{figure}

Our investigation is carried out through the use of the Multiphase DBB modeling framework, a new and flexible model used to simulate incompressible two-phase flow through and around deformable porous media \cite{Carrillo2020MDBB}. It consists of five volume averaged fluid and solid conservation equations that are coupled by spatially-dependent momentum exchange and capillary force terms. The model is composed of a fluid mass conservation equation,
\begin{equation} \label{eq:fluid_mass_conservation}
\begin{split}
\frac{\partial {\phi }_f}{\partial t}+\nabla \cdot{\boldsymbol{U}}_f=0 \ \ \ \ \ \ \ \ & 
\end{split}
\end{equation}

\noindent a fluid saturation conservation equation,
\begin{equation} \label{eq:saturation_equation}
\begin{split}
\frac{\partial {\phi }_f{\alpha }_w}{\partial t}+\nabla \cdot\left({\alpha }_w{\boldsymbol{U}}_f\right)+\nabla \cdot\left({\phi }_f{\alpha }_w{\alpha }_n{\boldsymbol{U}}_r\right)&=0  
\end{split}
\end{equation}

\noindent a fluid momentum conservation equation,
\begin{equation} \label{eq:fluid_mom_conservation}
    \begin{split}
       \frac{\partial {\rho }_f{\boldsymbol{U}}_f}{\partial t}+\nabla \cdot\left(\frac{{\rho }_f}{{\phi }_f}{\boldsymbol{U}}_f{\boldsymbol{U}}_f\right)=-{\phi }_f\nabla p+{\phi }_f{\rho }_f\boldsymbol{g}+ \ \ & \\ \nabla  \cdot \boldsymbol{S} -{\phi }_f{\mu }k^{-1}\left({\boldsymbol{U}}_f-{\boldsymbol{U}}_s\right)+{\phi }_f{\boldsymbol{F}}_{c}-{\phi }_fp_c\nabla \alpha_w & 
    \end{split}
\end{equation}

\noindent a solid mass conservation equation,
\begin{equation} \label{eq:solid_mass_cons}
    \begin{split}
        \frac{\partial {\phi }_s}{\partial t}+\nabla \cdot\left({\phi }_s{\boldsymbol{U}}_s\right)=0 \ \ \ \ \  & 
    \end{split}
\end{equation}

\noindent and a solid momentum conservation equation,
\begin{equation} \label{eq:solid_mom_cons}
    \begin{split}
         - \nabla  \cdot \boldsymbol{\sigma }=&-{\phi }_s\nabla p +{{\phi }_s\rho }_s\boldsymbol{g} + \\ &{\phi }_f{\mu}k^{-1}\left({\boldsymbol{U}}_f-{\boldsymbol{U}}_s\right)-{\phi }_f{\boldsymbol{F}}_{c}-{\phi }_sp_c\nabla \alpha_w \ 
    \end{split}
\end{equation}

\noindent In the previous equations, ${\phi }_f$ is the fluid volume fraction, ${\phi }_s$ is the solid volume fraction, ${\alpha}_w$ is the wetting fluid saturation, ${\alpha}_n$ is the non-wetting fluid saturation, ${\boldsymbol{U}}_f$ is the single-field fluid velocity, ${\boldsymbol{U}}_s$ is the solid velocity, ${\boldsymbol{U}}_r$ is the relative velocity of the two immiscible fluids, $p$ is the single-field fluid pressure, $\boldsymbol{S}$ is the volume averaged fluid viscous stress tensor, $\boldsymbol{\sigma}$ is the volume averaged solid stress tensor, $\mu k^{-1}$ is the drag coefficient (a function of permeability $k$ and single-field fluid viscosity ${\mu}$), $\rho_s$ is the solid density, $\boldsymbol{g}$ is gravity, $p_c$ is the capillary pressure, and ${\boldsymbol{F}}_{c}$ represents additional capillary terms. Here, ``single-field" refers to averaged variables that depend on the properties of both fluids. Lastly, $\rho_f={\alpha}_w{\rho }_w+{\alpha}_n{\rho}_n$ and ${\mu}={\alpha }_w{\mu }_w+{\mu }_n{\rho }_n$ are the single-field fluid density and viscosity. The closed form representations for ${\boldsymbol{U}}_r,\ {\mu}k^{-1}$, $p_c$, and ${\boldsymbol{F}}_{c}$ can be found in the Supplemental Materials along with an in-depth description of the model.
 
As indicated in Fig. \ref{fig:conceptual}, the system of equations presented above asymptotically approaches the Navier-Stokes multiphase volume-of-fluid \cite{Hirt1981} equations in solid free regions (where ${\phi}_f=1$, $k$ is very large, and viscous drag is negligible) and multiphase Biot Theory in porous regions (where ${\phi }_f<1$, $k$ is small, $\textrm{Re}<1$, and drag dominates). This last point can be demonstrated by adding Eqs. \ref{eq:fluid_mom_conservation} and \ref{eq:solid_mom_cons} together within a porous domain, which results in the main governing equation used in Biot Theory \cite{Jha2014,Kim2013,Carrillo2020MDBB}:
\begin{equation} \label{Eq:Biot_theory} 
\nabla  \cdot \boldsymbol{\sigma }=\nabla p-(\phi_s\rho_s+\phi_f\rho_f)\boldsymbol{g}\boldsymbol{+}p_c\nabla {\alpha }_w 
\end{equation} 

A thorough discussion, derivation, and validation of this model can be found in Carrillo \& Bourg 2020 \cite{Carrillo2020MDBB} and related publications \cite{Carrillo2019a,Carrillo2020}. The two major limitation of the framework highlighted in these previous studies are as follows. First, there needs to be a clear length-scale separation between the averaging volume, the sub-REV heterogeneities, and the overall system \cite{Whitaker1986}. This condition is sustained in most situations involving fractured porous materials, where fracture width is generally significantly larger than the pore width within the porous matrix, with the possible exception of microfractures. Second, closure of the system of equations necessitates the use of parametric models describing the average behaviour of the capillary pressure, permeability, and solid rheology within porous domains. As such, the accuracy of the overall model is inherently impacted by the limitations and assumptions of these parametric models. The complete numerical implementation of the solver, its validations, and the cases shown within this study can be found within the open-source simulation package \href{https://github.com/Franjcf}{``hybridBiotInterFoam''} \cite{hybridBiotInterFoam_Code}.

\section{Numerical Simulations}

\subsection{Crossover from Imbibition to Fracturing in a Hele-Shaw Cell } \label{subsect:Huang}

\begin{figure}
\includegraphics[width=0.483\textwidth]{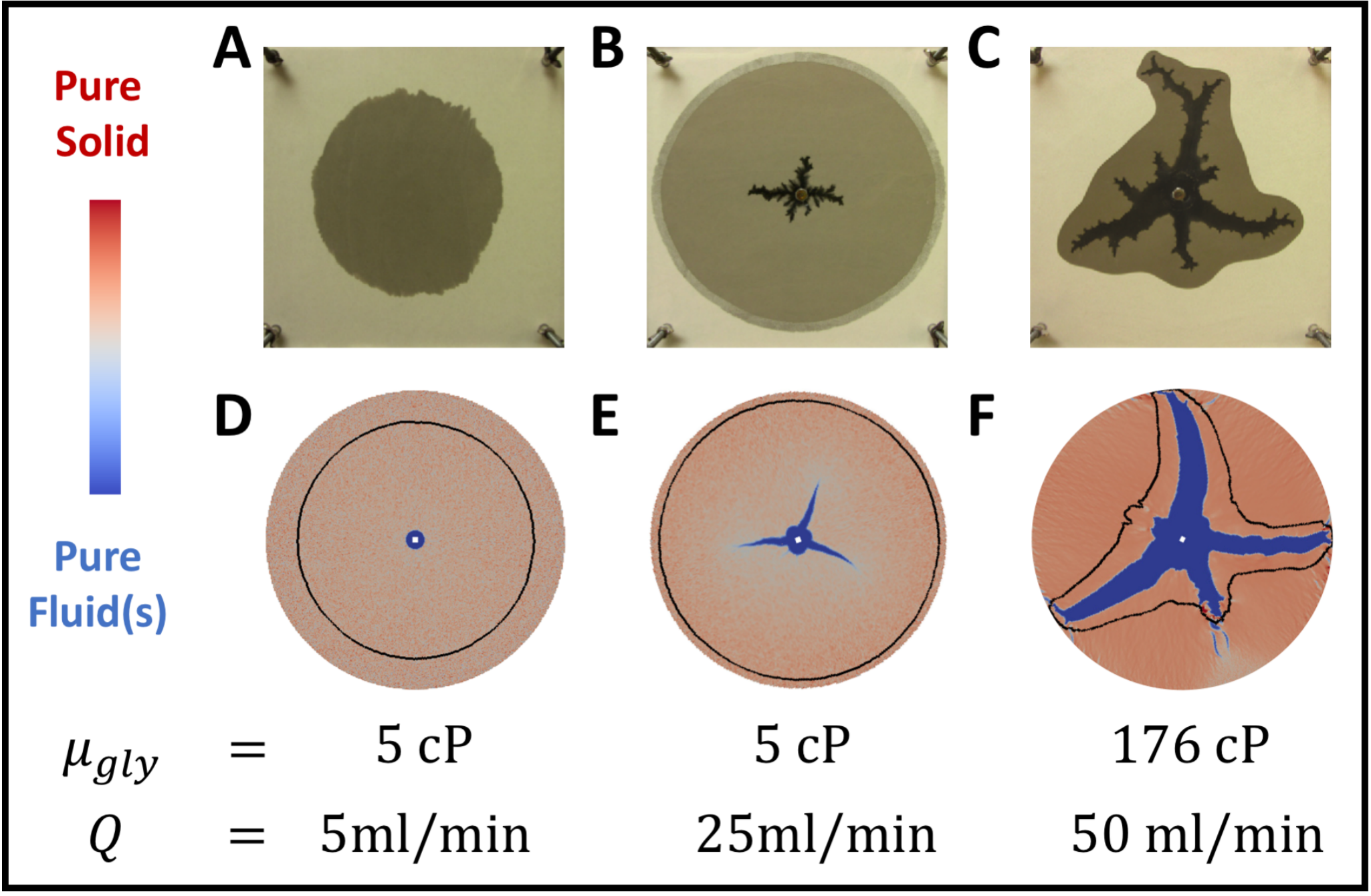}
\caption{\label{fig:Huang_Frac} Continuous transition from fluid imbibition to fracturing in a Hele-Shaw cell. Experimental images (A, B, C) were taken from Huang et. al. \cite{Huang2012a} and numerically replicated using equivalent conditions (D, E, F). Black lines represent the advancing saturation front. Additional cases can be found in \cite{Carrillo2020MDBB}.}
\end{figure}

In addition to the derivation and extensive quantitative validation of Eqs. \ref{eq:fluid_mass_conservation}-\ref{eq:solid_mom_cons}, our recent work \cite{Carrillo2020MDBB} included a qualitative validation of the ability of the Multiphase DBB model to predict the transition from invasion to fracturing during multiphase flow. Briefly, this validation replicated experiments by Huang et. al. \cite{Huang2012a} involving the injection of aqueous glycerin into dry sand at incremental flow rates within a 30 by 30 by 2.5 cm Hele-Shaw cell. As shown in Fig. \ref{fig:Huang_Frac}, these experiments are inherently multiphysics as fluid flow is governed by Stokes flow in the fracture (aperture $\sim$cm) and by multiphase Biot Theory in the porous sand (pore width $\sim 100\mu$m). As discussed in our previous work, the similarities between our model and the experimental results are evident: as the viscous forces imposed on the solid increase, so does the system’s propensity to exhibit fracturing as the primary flow mechanism (as opposed to imbibition). Minor microstructural differences between our simulations and the
experiments reflect the manner in which the implemented continuum-scale rheology model approximates the solid's granular nature. It is clear, however, that both systems are controlled by the balance between viscous forces and solid rheology at the scale of interest \cite{Carrillo2020MDBB}. As such, these experiments present an ideal starting point for our investigation.

\begin{figure*}
\includegraphics[width=0.982\textwidth]{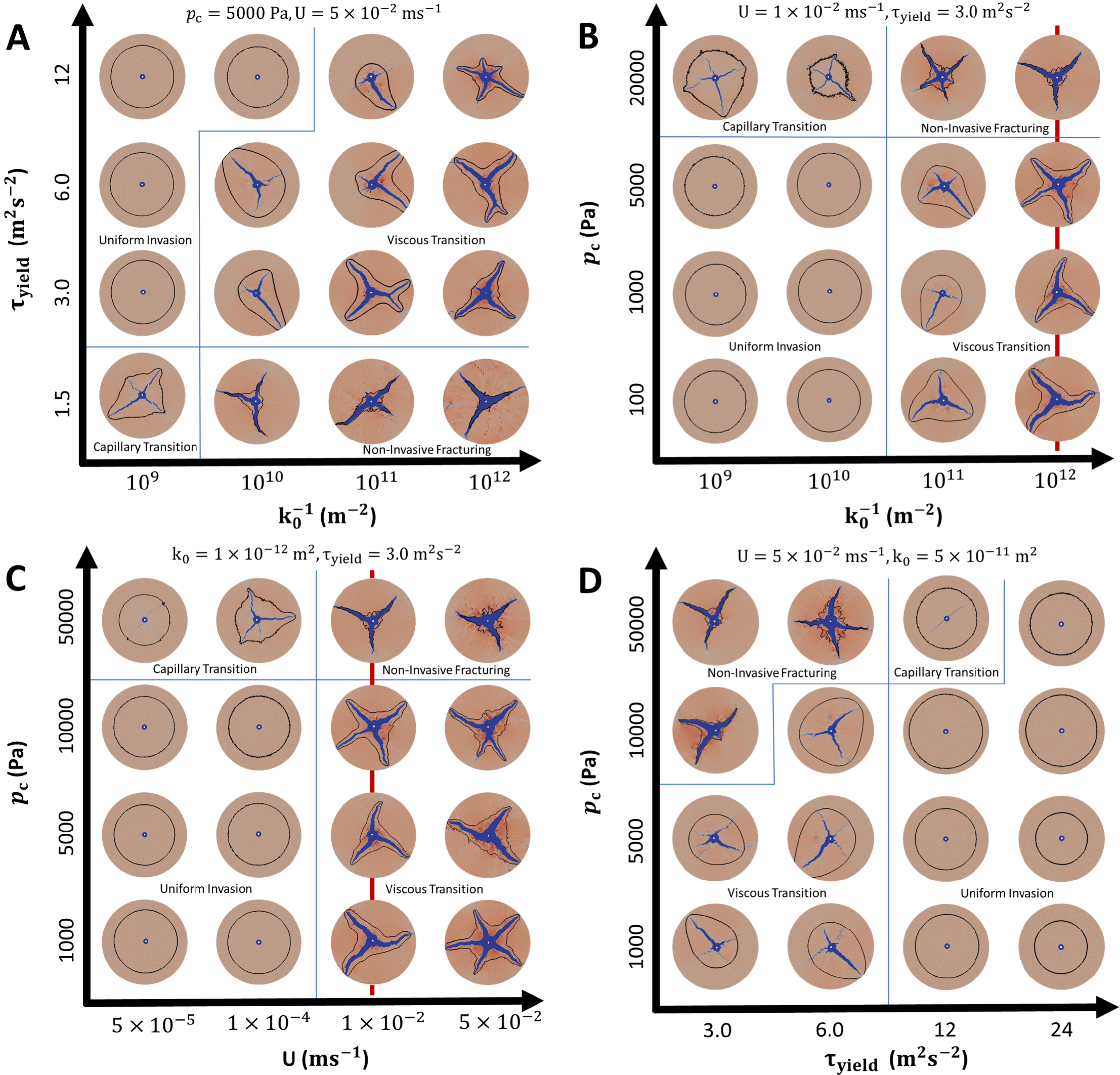}
\caption{\label{fig:phase_diagrams} Phase diagrams describing the effects of varying permeability, plastic yield stress, fluid injection rate, and capillary entry pressure on the transition from fluid drainage to fracturing. All cases are at ${\phi }_s=0.60\pm 0.05$ and ${\mu }_{n}=5\ \mathrm{cP}$. The remaining parameters are case-specific and can be found in each figure's upper legend. The areas separated by thin blue lines highlight and label the four deformation regimes described in Section \ref{sect:fracture_mechanisms}. The vertical red lines represent where these diagrams intersect in 3-dimensional space. The color scheme is the same as in Fig. \ref{fig:Huang_Frac}.}

\end{figure*}

\subsection{Creation of Fracturing Phase Diagrams}

Here, we use the same simulation methodology developed in \cite{Carrillo2020MDBB} and illustrated in Figure \ref{fig:Huang_Frac} to identify the general non-dimensional parameters that control the observed 
transitional behavior between invasion and fracturing in a plastic porous medium. To do so, we systematically vary
the solid's porosity ($\phi_f$ = 0.4 to 0.8), density-normalized plastic yield stress ($\tau_{yield}$ = $1.5$ to ${24\ \mathrm{m^2/s^2}}$), capillary entry pressure ($p_{c,0}$ = $100$ to $50,000\ \mathrm{Pa}$), and permeability ($k$ = $1\times10^{-13}$ and $5\times10^{-9} \
\mathrm{m^2}$) as well as the invading fluid's viscosity ($\mu_n$ = $0.5$ to $50\ \mathrm{cP}$) and injection rate ($\boldsymbol{U}_f$ = $1\times10^{-4}$ to $8\times10^{-2} \
\mathrm{m/s}$). As in our previous work, the solid's porosity was initialized as a normally-distributed field, the deformable solid was modeled as a Hershel-Bulkley-Quemada plastic \cite{Spearman2017,Quemada1977}, the porosity-dependence of permeability was modeled through the Kozeny-Carman relation, and relative permeabilities where calculated through the van-Genuchten model \cite{VanGenutchen1980}. Further details regarding the base numerical implementation of this model can be found in \cite{Carrillo2020MDBB}, the accompanying code \citep{hybridBiotInterFoam_Code}, and the Supplementary Materials. The only major differences relative to our previous simulations are that we now include capillary effects and
represent viscously-stable drainage as opposed to imbibition (i.e., the injected glycerin is now non-wetting to the porous medium). A representative sample of the more than 400 resulting simulations is presented in the phase
diagrams shown in Fig. \ref{fig:phase_diagrams}.

Overall, the results make intuitive sense. Figure
\ref{fig:phase_diagrams}A shows that, \textit{ceteris-paribus}, less
permeable solids are more prone to fracturing. This is due
to the fact that, given a constant flow rate, lower permeability solids experience greater drag forces. Our results also show that solids
with lower plastic yield stresses fracture more readily, as their solid structure cannot
withstand the effects of relatively large viscous or capillary forces. The y-axis behavior of
Fig. \ref{fig:phase_diagrams}B further shows that systems with higher entry pressures are more likely to
fracture, i.e., the capillary stresses are more likely to overwhelm the solid's yield stress, in agreement with grain scale simulations \cite{Jain2009}. Finally, Fig.
\ref{fig:phase_diagrams}B also shows that higher injection rates lead to more fracturing, as these
increase viscous drag on the solid structure.   

\section{Characterization of Fracturing Mechanisms} \label{sect:fracture_mechanisms}

The deformation regimes observed in the previous experiments can be
delineated by defining two simple non-dimensional parameters that quantify the balance
between viscous pressure drop, solid softness, and capillary entry pressure.

\begin{equation}\label{eq:N_vis_radial}
    N_{vF}=\frac{\Delta p}{{\tau }_{yield}{\rho }_s}=\frac{{\mu }U{r}_{in}}{k{\tau }_{yield}{\rho }_s}{\mathrm{ln} \left(\frac{r_{out}}{r_{in}}\right)}\ \ \ \  
\end{equation}

\begin{equation} \label{Eq:N_cap} 
N_{cF}=\frac{p_{c,0}}{{\tau }_{yield}{\rho }_s}=\frac{2\gamma }{r_{pore}{\tau }_{yield}{\rho }_s} 
\end{equation} 

\begin{figure}
\includegraphics[width=0.483\textwidth]{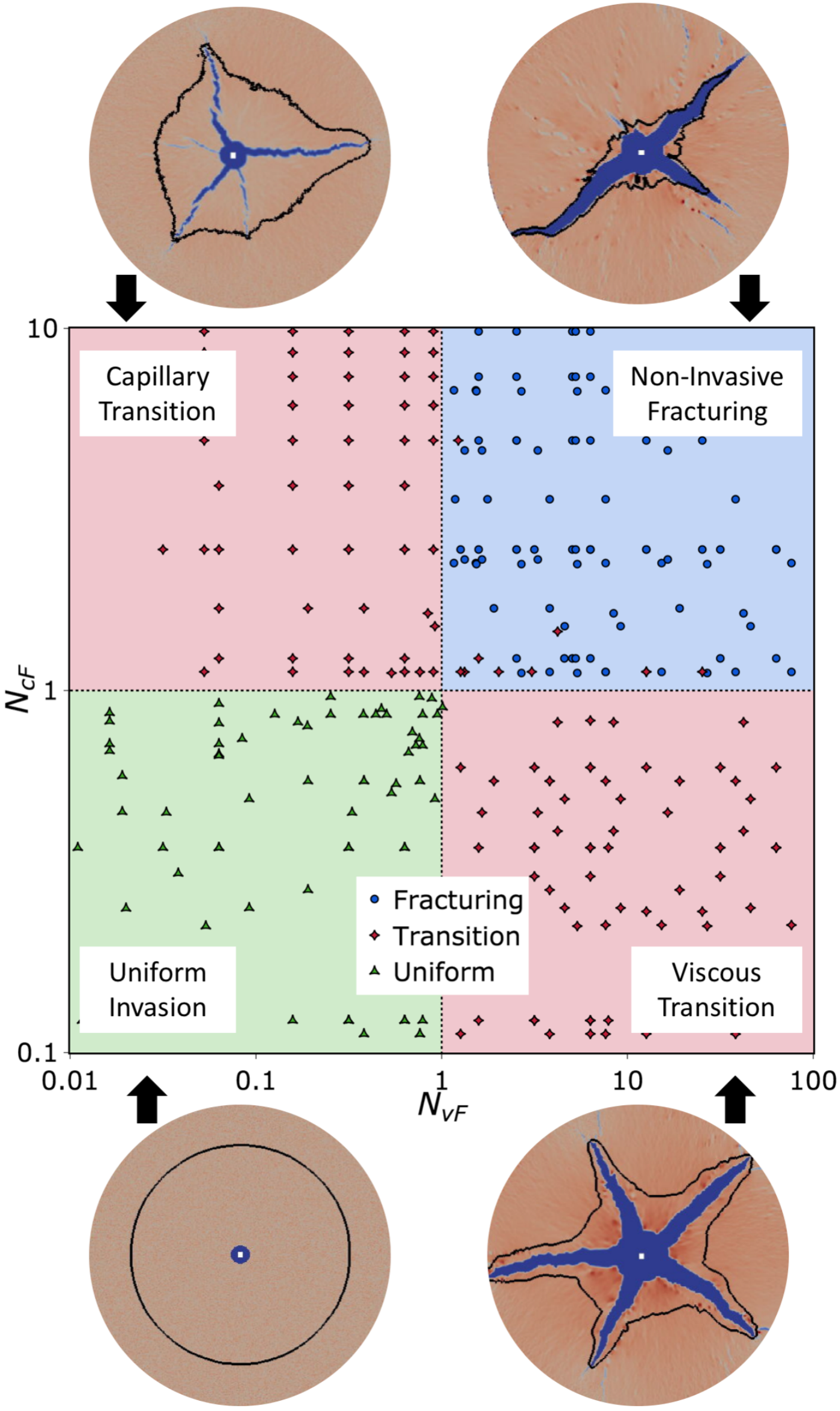}
\caption{\label{fig:Final_phase} Fluid invasion and fracturing in plastic porous media as a function of the viscous fracturing number $N_{vF}$ and the capillary fracturing number $N_{cF}$. Green triangles denote uniform invasion, red diamonds denote the transitional fracturing regimes, and blue circles denote non-invasive fracturing. The four images are representative samples of each fracturing regime.}
\end{figure}

Here, the viscous fracturing number ($N_{vF}$) represents the ratio between the viscous pressure drop and the solid's structural forces. It embodies the question: Does fluid flow generate sufficient friction to induce fracturing? As shown in Fig. \ref{fig:Final_phase}, the answer is no if $N_{vF}<1$ and yes if $N_{vF}>1.$ This number is the continuum scale analog to the fracturing number presented by Holtzman et. al. \cite{Holtzman2012a} for granular solids. It also explains the experimental finding by Zhou et. al. \cite{Zhou2010} that fracture initiation is only a function of the resulting fluid pressure drop, irrespective of the injection rate or fluid viscosity used to create it. Furthermore, it illustrates why increasing the injection rate and decreasing the permeability have similar effects in Fig. \ref{fig:phase_diagrams}.

Complementarily, the capillary fracturing number ($N_{cF}$) represents the ratio between the capillary entry pressure and the solid's structural forces; it embodies the question: Does multiphase flow generate sufficient capillary stresses to fracture the solid? Figure \ref{fig:Final_phase} shows that when $N_{cF}<1$ drainage is the preferential flow mechanism and when $N_{cF}>1$ fracturing becomes the dominant phenomenon.

This analysis yields the rudimentary conclusion that fracturing should occur if either of the fracturing numbers is greater than unity, as confirmed by our simulations. However, our simulations further demonstrate the existence of three distinct fracturing regimes (Figs. \ref{fig:phase_diagrams}-\ref{fig:Final_phase}). The first regime, referred here as \textit{non-invasive fracturing} ($N_{vF}>1$ and $N_{cF}>1$) is characterized by fracturing of the porous solid with minimal fluid invasion, where fractures precede any invasion front. In the second regime, referred to here as the \textit{viscous fracturing transition} ($N_{vF}>1$ and $N_{cF}<1$), only the viscous stresses are sufficiently large to fracture the solid. This leads to the formation of relatively wide fractures enveloped and preceded by a non-uniform invasion front. Finally, in the third regime, referred to here as the \textit{capillary fracturing transition} ($N_{vF}<1$ and $N_{cF}>1$), only the capillary stresses are sufficiently large to fracture the solid. Given a constant injection rate, this leads to the formation of fractures preceded by an invasion front, as in the viscous fracturing transition regime, but with a more uniform saturation front (due to lower viscous stresses) and less solid compaction (hence narrower fractures). We note that the crossover between each of the four regimes is continuous, meaning that systems with $N_{vF}$ or $N_{cF}\sim1$ can share elements of neighboring regimes. 

\begin{figure*}[t!]
\includegraphics[width=0.983\textwidth]{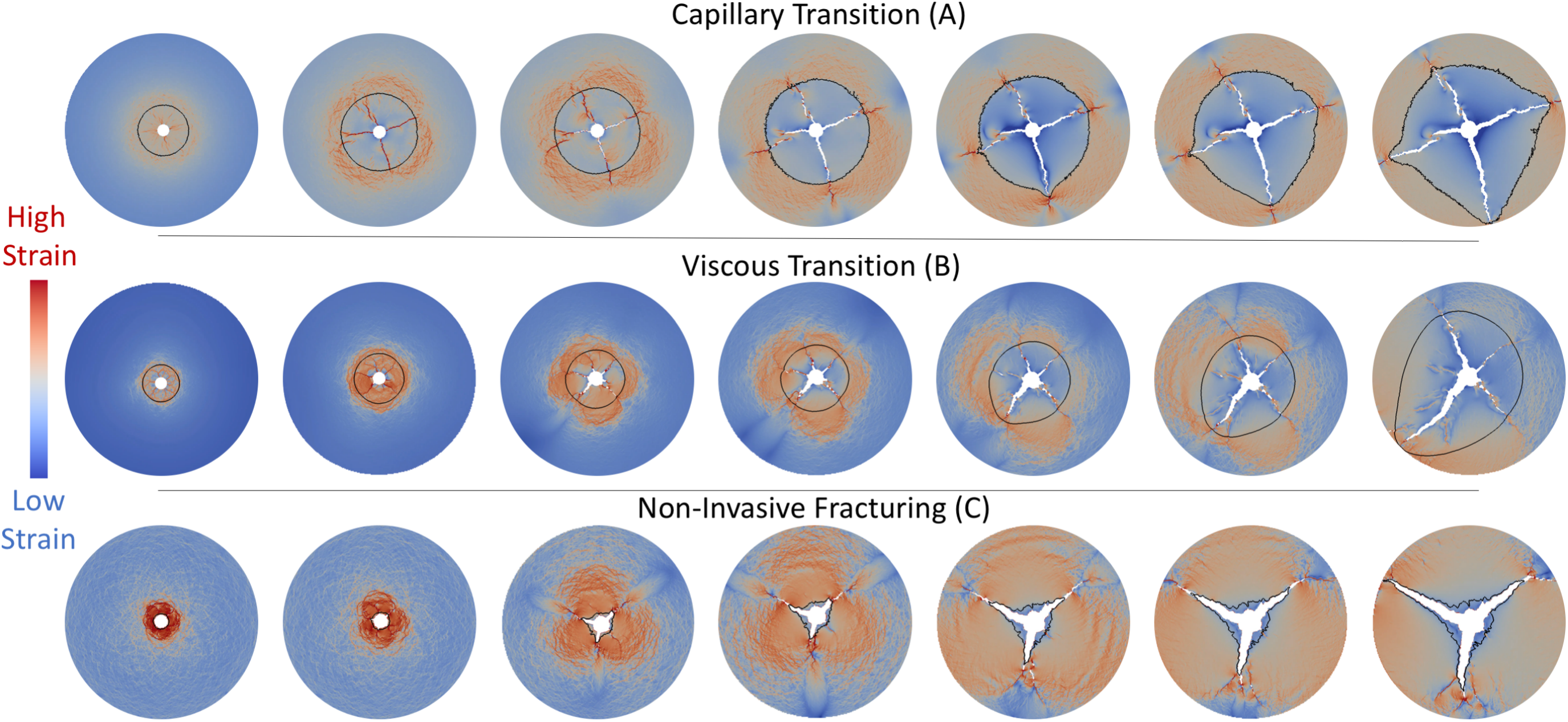}
\caption{\label{fig:strain} Dynamic fracture formation mechanisms. Each row represents the time-dependant fracture formation process for each fracturing type, where time advances from left to right. Here, the red-blue color scheme represents the log-normalized strain-rate magnitude specific to each simulated case, fractures are shown in white and the advancing fluid-fluid interface is shown as a thin black line.}
\end{figure*}

Although $N_{vF}$ and $N_{cF}$ are fairly intuitive numbers, their impacts on fracture propagation mechanisms are not. For this reason, we also studied the dynamics of fracture nucleation and growth and the evolution of the solid's strain for all three fracturing regimes. As seen in Fig. \ref{fig:strain}, fracturing in the two transition zones is characterized by the initial formation of non-flow-bearing failure zones (hereafter referred to as cracks), which function as nucleation sites for propagating flow-bearing fractures. These cracks are formed by the simultaneous movement of large contiguous sections of the porous medium in different directions, a process induced by uniform fluid invasion into the porous medium. However, the similarities between both transition zones end here. In the viscous fracturing transition regime, fractures quickly become the dominant deformation mechanism, localizing the majority of the stresses and solid compaction around the advancing fracture tip. Conversely, in the capillary fracturing transition regime, fluid-invasion continues to serve as the main flow mechanism and source of deformation, where fractures and cracks are slowly expanded due to the more evenly-distributed capillarity-induced stresses localized at the advancing invasion front. Finally, non-invasive fracturing follows a different process, where there is little-to-no crack formation and fracture propagation is the main source of deformation and flow. Here, the co-advancing fracture and saturation fronts uniformly compress the solid around and in front of them until this deformation reaches the outer boundary of the simulated system (see the ``jet" like-structures at fracture tips in Fig. \ref{fig:strain}C.) Pressure profiles that further showcase these behaviours can be found in the Supplementary Materials.  

\section{Influence of Localized and Uniform Deformation}

So far we have explored how independently changing $k$, $p_c$, and $\tau_{yield}$ (among others) can affect the fracturing of plastic materials. However, our results also have implications for situations in which these variables are all varied simultaneously, such as during the compaction of soils, sediments, or viscoplastic sedimentary rocks (i.e. mudstones or clay-shales). In such situations, with increasing compaction, $k^{-1}$, $p_c$, and $\tau_{yield}$ should all increase, although at different rates. As such, we now study the effects of local and uniform deformation on the outlined fracturing regimes. 

\subsection{Localized Deformation}

The simulations presented above were carried out using the simplifying assumption that $p_c$ is invariant with $\phi_f$ (whereas $k$ and $\tau_{yield}$ are not). To evaluate the impact of this simplification on the results shown in Figs. \ref{fig:phase_diagrams}-\ref{fig:Final_phase}, we carried out additional simulations for all four regimes with a deformation-dependent capillary entry pressure based on a simplified form of the Leverett J-function where $p_{c,0} = p_{c,0}^*(\phi_s/\phi_s^{avg})^n$, $p_{c,0}^*$ is the capillary pressure at $\phi_s=\phi_s^{avg}$, and $n>0$ is a sensitivity parameter \cite{CLeverett,Li2015}. The results show that non-zero values of $n$ promote the creation of finger-like instabilities and the nucleation of cracks at the fluid invasion front, particularly in the capillary fracturing transition regime. Simulation predictions with different \textit{n} values are shown in Fig. \ref{fig:variable_pc} in the capillary fracturing transition regime and in Supplemental Materials in other regimes. 

Despite the additional complexity of the resulting fluid invasion and fracturing patterns, results with $n>0$ conform to the overall phase diagram presented in Fig. \ref{fig:Final_phase}. The results at $n=0$ are therefore highlighted in the previous sections due to the greater simplicity of their fluid and solid distribution patterns.

\begin{figure}[b!] 
\includegraphics[width=0.48\textwidth]{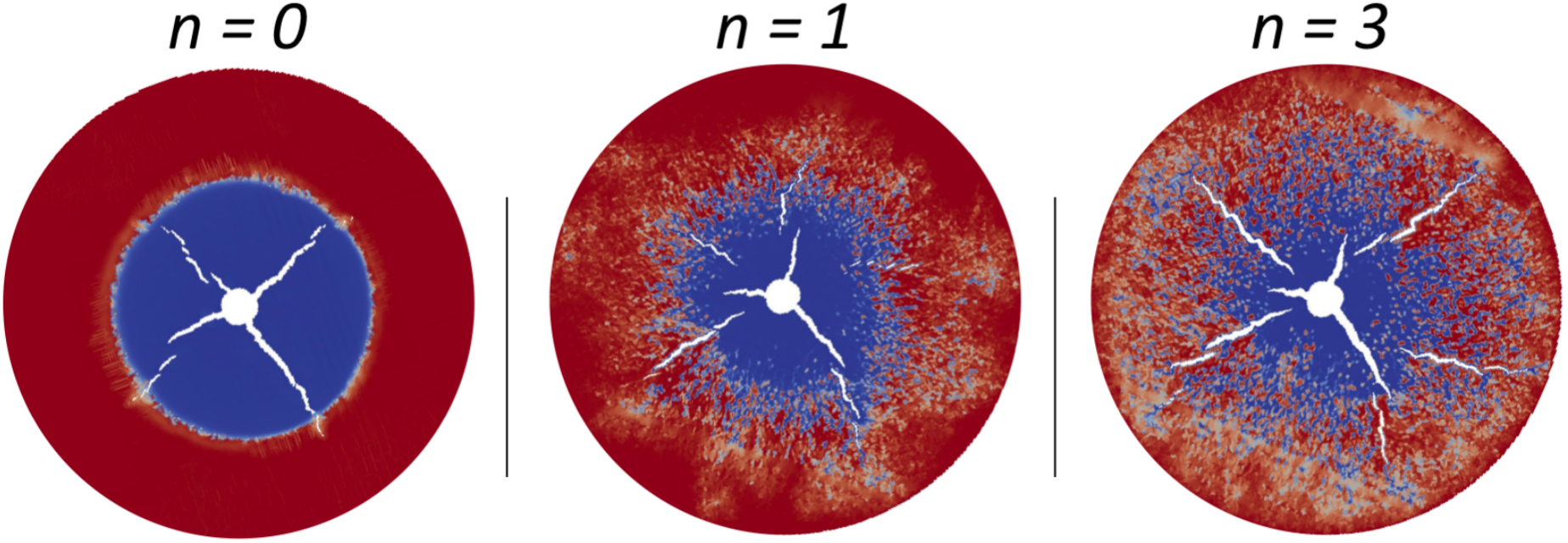}
\caption{\label{fig:variable_pc} Influence of the $\phi_f$-dependence of $p_c$ on fluid invasion (red and blue) and fracturing patterns (white) in the capillary fracturing transition regime. Here, $n$ represents the sensitivity parameter in the Leverett J-function analogue presented above.}
\end{figure}

\subsection{Uniform Deformation}

Having verified that the applicability of the fracturing numbers holds for systems were $k$, $\tau_{yield}$, and $p_c$ all vary with $\phi_f$, we now examine the effects of uniform compaction on said numbers. A direct analysis using the widely-used porosity-parameter relationships implemented above (the Kozeny-Carman relation for $k$, Leverett J-Function for $p_c$, and Quemada model for $\tau_{yield}$ \cite{CLeverett,Quemada1977,Spearman2017}) yields the following fracturing number - porosity dependence:

\begin{equation}\label{eq:N_vis_compaction}
    N_{vF} \propto \frac{(1-\phi_f)^{2-D}(1-\phi_{f,min}/\phi_f)}{\phi_f^2} 
\end{equation}

\begin{equation} \label{Eq:N_cap_compaction} 
N_{cF} \propto (1-\phi_f)^{2-D}(1-\phi_{f,min}/\phi_f)  
\end{equation} 

\noindent where $D$ is a rheological parameter based on the solid's fractal dimension (common values range for 1.7-2.9 for different clayey sediments \cite{Spearman2017}) and $\phi_{f,min}$ is the maximum possible degree of compaction. Through these relations, we can see that uniform compaction (or expansion) has a highly non-linear effect on fracturing. Equations \ref{eq:N_vis_compaction}-\ref{Eq:N_cap_compaction} indicate that whereas $N_{cF}$ tends to consistently decrease with increasing compaction, $N_{vF}$ is considerably more susceptible to changes in $\phi_f$ and exhibits multiple changes in the sign of its first derivative when $D > 2$, non-intuitively suggesting that fracturing can be either induced or suppressed through uniform compression. Plots of $N_{vF}$ and $N_{cF}$ as a function of solid fraction are reported in the Supplementary Materials.

\section{Conclusions}

In this article, we used the Multiphase DBB modeling framework to create a phase diagram that identifies two non-dimensional parameters that categorize the crossover between viscously-stable fluid drainage and fracturing as a function of wettability, solid deformability, and hydrodynamics. To the best of our knowledge, our results are the first to relate all three of these properties to characterize multiphase flow in viscoplastic porous media. As expected intuitively, we observe that fracturing occurs if the viscous and/or capillary stresses are sufficient to overcome the solid's structural forces. Thus, when it comes to systems with multiple fluids, it is necessary to consider the effects of surface tension, wettability, and pore size on the fluids' propensity to fracture or invade the permeable solid. Lastly, we find that the two non-dimensional fracturing numbers identified above delineate the existence of three fracturing regimes with distinct fracture propagation mechanisms.

\begin{acknowledgments}
This work was supported by the National Science Foundation, Division of Earth Sciences, Early Career program through Award EAR-1752982. FJC acknowledges additional support from the Mary and Randall Hack ‘69 Fellowship of the High Meadows Environmental Institute at Princeton University.
\end{acknowledgments}

\appendix

\bibliography{ref,SIbib}%

\begin{thebibliography}{39}%
\makeatletter
\providecommand \@ifxundefined [1]{%
 \@ifx{#1\undefined}
}%
\providecommand \@ifnum [1]{%
 \ifnum #1\expandafter \@firstoftwo
 \else \expandafter \@secondoftwo
 \fi
}%
\providecommand \@ifx [1]{%
 \ifx #1\expandafter \@firstoftwo
 \else \expandafter \@secondoftwo
 \fi
}%
\providecommand \natexlab [1]{#1}%
\providecommand \enquote  [1]{``#1''}%
\providecommand \bibnamefont  [1]{#1}%
\providecommand \bibfnamefont [1]{#1}%
\providecommand \citenamefont [1]{#1}%
\providecommand \href@noop [0]{\@secondoftwo}%
\providecommand \href [0]{\begingroup \@sanitize@url \@href}%
\providecommand \@href[1]{\@@startlink{#1}\@@href}%
\providecommand \@@href[1]{\endgroup#1\@@endlink}%
\providecommand \@sanitize@url [0]{\catcode `\\12\catcode `\$12\catcode
  `\&12\catcode `\#12\catcode `\^12\catcode `\_12\catcode `\%12\relax}%
\providecommand \@@startlink[1]{}%
\providecommand \@@endlink[0]{}%
\providecommand \url  [0]{\begingroup\@sanitize@url \@url }%
\providecommand \@url [1]{\endgroup\@href {#1}{\urlprefix }}%
\providecommand \urlprefix  [0]{URL }%
\providecommand \Eprint [0]{\href }%
\providecommand \doibase [0]{https://doi.org/}%
\providecommand \selectlanguage [0]{\@gobble}%
\providecommand \bibinfo  [0]{\@secondoftwo}%
\providecommand \bibfield  [0]{\@secondoftwo}%
\providecommand \translation [1]{[#1]}%
\providecommand \BibitemOpen [0]{}%
\providecommand \bibitemStop [0]{}%
\providecommand \bibitemNoStop [0]{.\EOS\space}%
\providecommand \EOS [0]{\spacefactor3000\relax}%
\providecommand \BibitemShut  [1]{\csname bibitem#1\endcsname}%
\let\auto@bib@innerbib\@empty
\bibitem [{\citenamefont {B{\"{a}}cher}\ and\ \citenamefont
  {Gekle}(2019)}]{Bacher2019}%
  \BibitemOpen
  \bibfield  {author} {\bibinfo {author} {\bibfnamefont {C.}~\bibnamefont
  {B{\"{a}}cher}}\ and\ \bibinfo {author} {\bibfnamefont {S.}~\bibnamefont
  {Gekle}},\ }\bibfield  {title} {\bibinfo {title} {{Computational modeling of
  active deformable membranes embedded in three-dimensional flows}},\ }\href
  {https://doi.org/10.1103/PhysRevE.99.062418} {\bibfield  {journal} {\bibinfo
  {journal} {Physical Review E}\ }\textbf {\bibinfo {volume} {99}},\ \bibinfo
  {pages} {062418} (\bibinfo {year} {2019})}\BibitemShut {NoStop}%
\bibitem [{\citenamefont {R{\"{a}}ss}\ \emph {et~al.}(2018)\citenamefont
  {R{\"{a}}ss}, \citenamefont {Simon},\ and\ \citenamefont
  {Podladchikov}}]{Rass2018}%
  \BibitemOpen
  \bibfield  {author} {\bibinfo {author} {\bibfnamefont {L.}~\bibnamefont
  {R{\"{a}}ss}}, \bibinfo {author} {\bibfnamefont {N.~S.}\ \bibnamefont
  {Simon}},\ and\ \bibinfo {author} {\bibfnamefont {Y.~Y.}\ \bibnamefont
  {Podladchikov}},\ }\bibfield  {title} {\bibinfo {title} {{Spontaneous
  formation of fluid escape pipes from subsurface reservoirs}},\ }\href
  {https://doi.org/10.1038/s41598-018-29485-5} {\bibfield  {journal} {\bibinfo
  {journal} {Scientific Reports}\ }\textbf {\bibinfo {volume} {8}},\ \bibinfo
  {pages} {11116} (\bibinfo {year} {2018})}\BibitemShut {NoStop}%
\bibitem [{\citenamefont {Towner}(1987)}]{Towner1987}%
  \BibitemOpen
  \bibfield  {author} {\bibinfo {author} {\bibfnamefont {G.~D.}\ \bibnamefont
  {Towner}},\ }\bibfield  {title} {\bibinfo {title} {{The mechanics of cracking
  of drying clay}},\ }\href {https://doi.org/10.1016/0021-8634(87)90118-1}
  {\bibfield  {journal} {\bibinfo  {journal} {Journal of Agricultural
  Engineering Research}\ }\textbf {\bibinfo {volume} {36}},\ \bibinfo {pages}
  {115} (\bibinfo {year} {1987})}\BibitemShut {NoStop}%
\bibitem [{\citenamefont {Lenormand}(1986)}]{Lenormand1986}%
  \BibitemOpen
  \bibfield  {author} {\bibinfo {author} {\bibfnamefont {R.}~\bibnamefont
  {Lenormand}},\ }\bibfield  {title} {\bibinfo {title} {{Pattern growth and
  fluid displacements through porous media}},\ }\href
  {https://doi.org/10.1016/0378-4371(86)90211-6} {\bibfield  {journal}
  {\bibinfo  {journal} {Physica A: Statistical Mechanics and its Applications}\
  }\textbf {\bibinfo {volume} {140}},\ \bibinfo {pages} {114} (\bibinfo {year}
  {1986})}\BibitemShut {NoStop}%
\bibitem [{\citenamefont {M{\aa}l{\o}y}\ \emph {et~al.}(1985)\citenamefont
  {M{\aa}l{\o}y}, \citenamefont {Feder},\ and\ \citenamefont
  {J{\o}ssang}}]{JorgenMalby1985}%
  \BibitemOpen
  \bibfield  {author} {\bibinfo {author} {\bibfnamefont {K.~J.}\ \bibnamefont
  {M{\aa}l{\o}y}}, \bibinfo {author} {\bibfnamefont {J.}~\bibnamefont
  {Feder}},\ and\ \bibinfo {author} {\bibfnamefont {T.}~\bibnamefont
  {J{\o}ssang}},\ }\bibfield  {title} {\bibinfo {title} {{Viscous Fingering
  Fractals in Porous Media}},\ }\href
  {https://journals.aps.org/prl/pdf/10.1103/PhysRevLett.55.2688} {\bibfield
  {journal} {\bibinfo  {journal} {Physical Review Letters}\ }\textbf {\bibinfo
  {volume} {55}},\ \bibinfo {pages} {2688} (\bibinfo {year}
  {1985})}\BibitemShut {NoStop}%
\bibitem [{\citenamefont {Saffman}\ and\ \citenamefont
  {Taylor}(1958)}]{Saffman_1958}%
  \BibitemOpen
  \bibfield  {author} {\bibinfo {author} {\bibfnamefont {P.~G.}\ \bibnamefont
  {Saffman}}\ and\ \bibinfo {author} {\bibfnamefont {G.~I.}\ \bibnamefont
  {Taylor}},\ }\bibfield  {title} {\bibinfo {title} {{The penetration of a
  fluid into a porous medium or Hele-Shaw cell containing a more viscous
  liquid}},\ }\href {https://doi.org/https://doi.org/10.1098/rspa.1958.0085}
  {\bibfield  {journal} {\bibinfo  {journal} {Proceedings of the Royal Society
  A}\ }\textbf {\bibinfo {volume} {245}},\ \bibinfo {pages} {312} (\bibinfo
  {year} {1958})}\BibitemShut {NoStop}%
\bibitem [{\citenamefont {Stokes}\ \emph {et~al.}(1986)\citenamefont {Stokes},
  \citenamefont {Weitz}, \citenamefont {Gollub}, \citenamefont {Dougherty},
  \citenamefont {Robbins}, \citenamefont {Chaikin},\ and\ \citenamefont
  {Lindsay}}]{Stokes1976}%
  \BibitemOpen
  \bibfield  {author} {\bibinfo {author} {\bibfnamefont {J.~P.}\ \bibnamefont
  {Stokes}}, \bibinfo {author} {\bibfnamefont {D.~A.}\ \bibnamefont {Weitz}},
  \bibinfo {author} {\bibfnamefont {J.~P.}\ \bibnamefont {Gollub}}, \bibinfo
  {author} {\bibfnamefont {A.}~\bibnamefont {Dougherty}}, \bibinfo {author}
  {\bibfnamefont {M.~O.}\ \bibnamefont {Robbins}}, \bibinfo {author}
  {\bibfnamefont {P.~M.}\ \bibnamefont {Chaikin}},\ and\ \bibinfo {author}
  {\bibfnamefont {H.~M.}\ \bibnamefont {Lindsay}},\ }\bibfield  {title}
  {\bibinfo {title} {{Interfacial stability of immiscible displacement in a
  porous medium}},\ }\href {https://doi.org/10.1103/PhysRevLett.57.1718}
  {\bibfield  {journal} {\bibinfo  {journal} {Physical Review Letters}\
  }\textbf {\bibinfo {volume} {57}},\ \bibinfo {pages} {1718} (\bibinfo {year}
  {1986})}\BibitemShut {NoStop}%
\bibitem [{\citenamefont {Ferer}\ \emph {et~al.}(2004)\citenamefont {Ferer},
  \citenamefont {Ji}, \citenamefont {Bromhal}, \citenamefont {Cook},
  \citenamefont {Ahmadi},\ and\ \citenamefont {Smith}}]{Ferer2004}%
  \BibitemOpen
  \bibfield  {author} {\bibinfo {author} {\bibfnamefont {M.}~\bibnamefont
  {Ferer}}, \bibinfo {author} {\bibfnamefont {C.}~\bibnamefont {Ji}}, \bibinfo
  {author} {\bibfnamefont {G.~S.}\ \bibnamefont {Bromhal}}, \bibinfo {author}
  {\bibfnamefont {J.}~\bibnamefont {Cook}}, \bibinfo {author} {\bibfnamefont
  {G.}~\bibnamefont {Ahmadi}},\ and\ \bibinfo {author} {\bibfnamefont {D.~H.}\
  \bibnamefont {Smith}},\ }\bibfield  {title} {\bibinfo {title} {{Crossover
  from capillary fingering to viscous fingering for immiscible unstable flow:
  Experiment and modeling}},\ }\href
  {https://doi.org/10.1103/PhysRevE.70.016303} {\bibfield  {journal} {\bibinfo
  {journal} {Physical Review E}\ }\textbf {\bibinfo {volume} {70}},\ \bibinfo
  {pages} {016303} (\bibinfo {year} {2004})}\BibitemShut {NoStop}%
\bibitem [{\citenamefont {Yortsos}\ \emph {et~al.}(1997)\citenamefont
  {Yortsos}, \citenamefont {Xu},\ and\ \citenamefont {Salin}}]{Yortsos1997}%
  \BibitemOpen
  \bibfield  {author} {\bibinfo {author} {\bibfnamefont {Y.~C.}\ \bibnamefont
  {Yortsos}}, \bibinfo {author} {\bibfnamefont {B.}~\bibnamefont {Xu}},\ and\
  \bibinfo {author} {\bibfnamefont {D.}~\bibnamefont {Salin}},\ }\bibfield
  {title} {\bibinfo {title} {{Phase Diagram of Fully Developed Drainage in
  Porous Media}},\ }\href
  {https://journals.aps.org/prl/pdf/10.1103/PhysRevLett.79.4581} {\bibfield
  {journal} {\bibinfo  {journal} {Physical Review Letters}\ }\textbf {\bibinfo
  {volume} {79}},\ \bibinfo {pages} {4581} (\bibinfo {year}
  {1997})}\BibitemShut {NoStop}%
\bibitem [{\citenamefont {Picchi}\ and\ \citenamefont
  {Battiato}(2018)}]{Picchi2018}%
  \BibitemOpen
  \bibfield  {author} {\bibinfo {author} {\bibfnamefont {D.}~\bibnamefont
  {Picchi}}\ and\ \bibinfo {author} {\bibfnamefont {I.}~\bibnamefont
  {Battiato}},\ }\bibfield  {title} {\bibinfo {title} {{The Impact of
  Pore-Scale Flow Regimes on Upscaling of Immiscible Two-Phase Flow in Porous
  Media}},\ }\href {https://doi.org/10.1029/2018WR023172} {\bibfield  {journal}
  {\bibinfo  {journal} {Water Resources Research}\ }\textbf {\bibinfo {volume}
  {54}},\ \bibinfo {pages} {6683} (\bibinfo {year} {2018})}\BibitemShut
  {NoStop}%
\bibitem [{\citenamefont {Picchi}\ and\ \citenamefont
  {Battiato}(2019)}]{Picchi2019}%
  \BibitemOpen
  \bibfield  {author} {\bibinfo {author} {\bibfnamefont {D.}~\bibnamefont
  {Picchi}}\ and\ \bibinfo {author} {\bibfnamefont {I.}~\bibnamefont
  {Battiato}},\ }\bibfield  {title} {\bibinfo {title} {{Relative Permeability
  Scaling From Pore-Scale Flow Regimes}},\ }\href
  {https://doi.org/10.1029/2018WR024251} {\bibfield  {journal} {\bibinfo
  {journal} {Water Resources Research}\ }\textbf {\bibinfo {volume} {55}},\
  \bibinfo {pages} {3215} (\bibinfo {year} {2019})}\BibitemShut {NoStop}%
\bibitem [{\citenamefont {Brooks}\ and\ \citenamefont
  {Corey}(1964)}]{RHBrooks1964}%
  \BibitemOpen
  \bibfield  {author} {\bibinfo {author} {\bibfnamefont {R.~H.}\ \bibnamefont
  {Brooks}}\ and\ \bibinfo {author} {\bibfnamefont {A.~T.}\ \bibnamefont
  {Corey}},\ }\bibfield  {title} {\bibinfo {title} {{Hydraulic Properties of
  Porous Media}},\ }\href {http://www.citeulike.org/group/1336/article/711012}
  {\bibfield  {journal} {\bibinfo  {journal} {Hydrology Papers, Colorado State
  University}\ }\textbf {\bibinfo {volume} {3}},\ \bibinfo {pages} {37 pp}
  (\bibinfo {year} {1964})}\BibitemShut {NoStop}%
\bibitem [{\citenamefont {van Genuchten}(1980)}]{VanGenutchen1980}%
  \BibitemOpen
  \bibfield  {author} {\bibinfo {author} {\bibfnamefont {M.~T.}\ \bibnamefont
  {van Genuchten}},\ }\bibfield  {title} {\bibinfo {title} {{A Closed-form
  Equation for Predicting the Hydraulic Conductivity of Unsaturated Soils}},\
  }\href {https://doi.org/10.2136/sssaj1980.03615995004400050002x} {\bibfield
  {journal} {\bibinfo  {journal} {Soil Science Society of America Journal}\
  }\textbf {\bibinfo {volume} {44}},\ \bibinfo {pages} {892} (\bibinfo {year}
  {1980})}\BibitemShut {NoStop}%
\bibitem [{\citenamefont {Biot}(1941)}]{Biot1941}%
  \BibitemOpen
  \bibfield  {author} {\bibinfo {author} {\bibfnamefont {M.~A.}\ \bibnamefont
  {Biot}},\ }\bibfield  {title} {\bibinfo {title} {{General Theory of
  Three-Dimensional Consolidation}},\ }\href
  {https://doi.org/10.1063/1.1712886} {\bibfield  {journal} {\bibinfo
  {journal} {Journal of Applied Physics}\ }\textbf {\bibinfo {volume} {12}},\
  \bibinfo {pages} {155} (\bibinfo {year} {1941})}\BibitemShut {NoStop}%
\bibitem [{\citenamefont {Terzaghi}(1943)}]{Terzaghi1943}%
  \BibitemOpen
  \bibfield  {author} {\bibinfo {author} {\bibfnamefont {K.}~\bibnamefont
  {Terzaghi}},\ }\href {https://doi.org/10.1002/9780470172766} {\emph {\bibinfo
  {title} {{Theoretical Soil Mechanics}}}}\ (\bibinfo  {publisher} {John Wiley
  {\&} Sons, Inc.},\ \bibinfo {address} {Hoboken, NJ, USA},\ \bibinfo {year}
  {1943})\BibitemShut {NoStop}%
\bibitem [{\citenamefont {Auton}\ and\ \citenamefont
  {MacMinn}(2017)}]{Auton2017a}%
  \BibitemOpen
  \bibfield  {author} {\bibinfo {author} {\bibfnamefont {L.~C.}\ \bibnamefont
  {Auton}}\ and\ \bibinfo {author} {\bibfnamefont {C.~W.}\ \bibnamefont
  {MacMinn}},\ }\bibfield  {title} {\bibinfo {title} {{From arteries to
  boreholes: steady-state response of a poroelastic cylinder to fluid
  injection}},\ }\href {https://doi.org/10.1098/rspa.2016.0753} {\bibfield
  {journal} {\bibinfo  {journal} {Proceedings of the Royal Society A}\ }\textbf
  {\bibinfo {volume} {473}},\ \bibinfo {pages} {20160753} (\bibinfo {year}
  {2017})}\BibitemShut {NoStop}%
\bibitem [{\citenamefont {Bertrand}\ \emph {et~al.}(2016)\citenamefont
  {Bertrand}, \citenamefont {Peixinho}, \citenamefont {Mukhopadhyay},\ and\
  \citenamefont {MacMinn}}]{Bertrand2016b}%
  \BibitemOpen
  \bibfield  {author} {\bibinfo {author} {\bibfnamefont {T.}~\bibnamefont
  {Bertrand}}, \bibinfo {author} {\bibfnamefont {J.}~\bibnamefont {Peixinho}},
  \bibinfo {author} {\bibfnamefont {S.}~\bibnamefont {Mukhopadhyay}},\ and\
  \bibinfo {author} {\bibfnamefont {C.~W.}\ \bibnamefont {MacMinn}},\
  }\bibfield  {title} {\bibinfo {title} {{Dynamics of Swelling and Drying in a
  Spherical Gel}},\ }\href {https://doi.org/10.1103/PhysRevApplied.6.064010}
  {\bibfield  {journal} {\bibinfo  {journal} {Physical Review Applied}\
  }\textbf {\bibinfo {volume} {6}},\ \bibinfo {pages} {064010} (\bibinfo {year}
  {2016})}\BibitemShut {NoStop}%
\bibitem [{\citenamefont {Carrillo}\ and\ \citenamefont
  {Bourg}(2019)}]{Carrillo2019a}%
  \BibitemOpen
  \bibfield  {author} {\bibinfo {author} {\bibfnamefont {F.~J.}\ \bibnamefont
  {Carrillo}}\ and\ \bibinfo {author} {\bibfnamefont {I.~C.}\ \bibnamefont
  {Bourg}},\ }\bibfield  {title} {\bibinfo {title} {{A Darcy-Brinkman-Biot
  Approach to Modeling the Hydrology and Mechanics of Porous Media Containing
  Macropores and Deformable Microporous Regions}},\ }\href
  {https://doi.org/10.1029/2019WR024712} {\bibfield  {journal} {\bibinfo
  {journal} {Water Resources Research}\ }\textbf {\bibinfo {volume} {55}},\
  \bibinfo {pages} {8096} (\bibinfo {year} {2019})}\BibitemShut {NoStop}%
\bibitem [{\citenamefont {MacMinn}\ \emph {et~al.}(2015)\citenamefont
  {MacMinn}, \citenamefont {Dufresne},\ and\ \citenamefont
  {Wettlaufer}}]{MacMinn2015a}%
  \BibitemOpen
  \bibfield  {author} {\bibinfo {author} {\bibfnamefont {C.~W.}\ \bibnamefont
  {MacMinn}}, \bibinfo {author} {\bibfnamefont {E.~R.}\ \bibnamefont
  {Dufresne}},\ and\ \bibinfo {author} {\bibfnamefont {J.~S.}\ \bibnamefont
  {Wettlaufer}},\ }\bibfield  {title} {\bibinfo {title} {{Fluid-driven
  deformation of a soft granular material}},\ }\href
  {https://doi.org/10.1103/PhysRevX.5.011020} {\bibfield  {journal} {\bibinfo
  {journal} {Physical Review X}\ }\textbf {\bibinfo {volume} {5}},\ \bibinfo
  {pages} {011020} (\bibinfo {year} {2015})}\BibitemShut {NoStop}%
\bibitem [{\citenamefont {Campbell}\ \emph {et~al.}(2017)\citenamefont
  {Campbell}, \citenamefont {Ozturk},\ and\ \citenamefont
  {Sandnes}}]{Campbell2017}%
  \BibitemOpen
  \bibfield  {author} {\bibinfo {author} {\bibfnamefont {J.~M.}\ \bibnamefont
  {Campbell}}, \bibinfo {author} {\bibfnamefont {D.}~\bibnamefont {Ozturk}},\
  and\ \bibinfo {author} {\bibfnamefont {B.}~\bibnamefont {Sandnes}},\
  }\bibfield  {title} {\bibinfo {title} {{Gas-Driven Fracturing of Saturated
  Granular Media}},\ }\href {https://doi.org/10.1103/PhysRevApplied.8.064029}
  {\bibfield  {journal} {\bibinfo  {journal} {Physical Review Applied}\
  }\textbf {\bibinfo {volume} {8}},\ \bibinfo {pages} {064029} (\bibinfo {year}
  {2017})}\BibitemShut {NoStop}%
\bibitem [{\citenamefont {Sandnes}\ \emph {et~al.}(2011)\citenamefont
  {Sandnes}, \citenamefont {Flekk{\o}y}, \citenamefont {Knudsen}, \citenamefont
  {M{\aa}l{\o}y},\ and\ \citenamefont {See}}]{Sandnes2011a}%
  \BibitemOpen
  \bibfield  {author} {\bibinfo {author} {\bibfnamefont {B.}~\bibnamefont
  {Sandnes}}, \bibinfo {author} {\bibfnamefont {E.~G.}\ \bibnamefont
  {Flekk{\o}y}}, \bibinfo {author} {\bibfnamefont {H.~A.}\ \bibnamefont
  {Knudsen}}, \bibinfo {author} {\bibfnamefont {K.~J.}\ \bibnamefont
  {M{\aa}l{\o}y}},\ and\ \bibinfo {author} {\bibfnamefont {H.}~\bibnamefont
  {See}},\ }\bibfield  {title} {\bibinfo {title} {{Patterns and flow in
  frictional fluid dynamics}},\ }\href {https://doi.org/10.1038/ncomms1289}
  {\bibfield  {journal} {\bibinfo  {journal} {Nature Communications}\ }\textbf
  {\bibinfo {volume} {2}},\ \bibinfo {pages} {288} (\bibinfo {year}
  {2011})}\BibitemShut {NoStop}%
\bibitem [{\citenamefont {Zhang}\ \emph {et~al.}(2013)\citenamefont {Zhang},
  \citenamefont {Damjanac},\ and\ \citenamefont {Huang}}]{Zhang2013}%
  \BibitemOpen
  \bibfield  {author} {\bibinfo {author} {\bibfnamefont {F.}~\bibnamefont
  {Zhang}}, \bibinfo {author} {\bibfnamefont {B.}~\bibnamefont {Damjanac}},\
  and\ \bibinfo {author} {\bibfnamefont {H.}~\bibnamefont {Huang}},\ }\bibfield
   {title} {\bibinfo {title} {{Coupled discrete element modeling of fluid
  injection into dense granular media}},\ }\href
  {https://doi.org/10.1002/jgrb.50204} {\bibfield  {journal} {\bibinfo
  {journal} {Journal of Geophysical Research: Solid Earth}\ }\textbf {\bibinfo
  {volume} {118}},\ \bibinfo {pages} {2703} (\bibinfo {year}
  {2013})}\BibitemShut {NoStop}%
\bibitem [{\citenamefont {Holtzman}\ and\ \citenamefont
  {Juanes}(2010)}]{Holtzman2010a}%
  \BibitemOpen
  \bibfield  {author} {\bibinfo {author} {\bibfnamefont {R.}~\bibnamefont
  {Holtzman}}\ and\ \bibinfo {author} {\bibfnamefont {R.}~\bibnamefont
  {Juanes}},\ }\bibfield  {title} {\bibinfo {title} {{Crossover from fingering
  to fracturing in deformable disordered media}},\ }\href@noop {} {\bibfield
  {journal} {\bibinfo  {journal} {Physical Review E}\ }\textbf {\bibinfo
  {volume} {82}} (\bibinfo {year} {2010})}\BibitemShut {NoStop}%
\bibitem [{\citenamefont {Holtzman}\ \emph {et~al.}(2012)\citenamefont
  {Holtzman}, \citenamefont {Szulczewski},\ and\ \citenamefont
  {Juanes}}]{Holtzman2012a}%
  \BibitemOpen
  \bibfield  {author} {\bibinfo {author} {\bibfnamefont {R.}~\bibnamefont
  {Holtzman}}, \bibinfo {author} {\bibfnamefont {M.~L.}\ \bibnamefont
  {Szulczewski}},\ and\ \bibinfo {author} {\bibfnamefont {R.}~\bibnamefont
  {Juanes}},\ }\bibfield  {title} {\bibinfo {title} {{Capillary fracturing in
  granular media}},\ }\href {https://doi.org/10.1103/PhysRevLett.108.264504}
  {\bibfield  {journal} {\bibinfo  {journal} {Physical Review Letters}\
  }\textbf {\bibinfo {volume} {108}},\ \bibinfo {pages} {264504} (\bibinfo
  {year} {2012})}\BibitemShut {NoStop}%
\bibitem [{\citenamefont {Jain}\ and\ \citenamefont {Juanes}(2009)}]{Jain2009}%
  \BibitemOpen
  \bibfield  {author} {\bibinfo {author} {\bibfnamefont {A.~K.}\ \bibnamefont
  {Jain}}\ and\ \bibinfo {author} {\bibfnamefont {R.}~\bibnamefont {Juanes}},\
  }\bibfield  {title} {\bibinfo {title} {{Preferential mode of gas invasion in
  sediments: Grain-scale mechanistic model of coupled multiphase fluid flow and
  sediment mechanics}},\ }\href {https://doi.org/10.1029/2008JB006002}
  {\bibfield  {journal} {\bibinfo  {journal} {Journal of Geophysical Research:
  Solid Earth}\ }\textbf {\bibinfo {volume} {114}},\ \bibinfo {pages} {B08101}
  (\bibinfo {year} {2009})}\BibitemShut {NoStop}%
\bibitem [{\citenamefont {Meng}\ \emph {et~al.}(2020)\citenamefont {Meng},
  \citenamefont {Primkulov}, \citenamefont {Yang}, \citenamefont {Kwok},\ and\
  \citenamefont {Juanes}}]{Meng2020}%
  \BibitemOpen
  \bibfield  {author} {\bibinfo {author} {\bibfnamefont {Y.}~\bibnamefont
  {Meng}}, \bibinfo {author} {\bibfnamefont {B.~K.}\ \bibnamefont {Primkulov}},
  \bibinfo {author} {\bibfnamefont {Z.}~\bibnamefont {Yang}}, \bibinfo {author}
  {\bibfnamefont {C.~Y.}\ \bibnamefont {Kwok}},\ and\ \bibinfo {author}
  {\bibfnamefont {R.}~\bibnamefont {Juanes}},\ }\bibfield  {title} {\bibinfo
  {title} {{Jamming transition and emergence of fracturing in wet granular
  media}},\ }\href {https://doi.org/10.1103/physrevresearch.2.022012}
  {\bibfield  {journal} {\bibinfo  {journal} {Physical Review Research}\
  }\textbf {\bibinfo {volume} {2}},\ \bibinfo {pages} {022012} (\bibinfo {year}
  {2020})}\BibitemShut {NoStop}%
\bibitem [{\citenamefont {Carrillo}\ and\ \citenamefont
  {Bourg}(2020{\natexlab{a}})}]{Carrillo2020MDBB}%
  \BibitemOpen
  \bibfield  {author} {\bibinfo {author} {\bibfnamefont {F.~J.}\ \bibnamefont
  {Carrillo}}\ and\ \bibinfo {author} {\bibfnamefont {I.~C.}\ \bibnamefont
  {Bourg}},\ }\bibfield  {title} {\bibinfo {title} {{Modeling Multiphase Flow
  Within and Around Deformable Porous Materials: A Darcy‐Brinkman‐Biot
  Approach}},\ }\href {https://doi.org/10.1029/2020wr028734} {\bibfield
  {journal} {\bibinfo  {journal} {Water Resources Research}\ }\textbf {\bibinfo
  {volume} {57}},\ \bibinfo {pages} {e2020WR028734} (\bibinfo {year}
  {2020}{\natexlab{a}})}\BibitemShut {NoStop}%
\bibitem [{\citenamefont {Hirt}\ and\ \citenamefont
  {Nichols}(1981)}]{Hirt1981}%
  \BibitemOpen
  \bibfield  {author} {\bibinfo {author} {\bibfnamefont {C.~W.}\ \bibnamefont
  {Hirt}}\ and\ \bibinfo {author} {\bibfnamefont {B.~D.}\ \bibnamefont
  {Nichols}},\ }\bibfield  {title} {\bibinfo {title} {{Volume of Fluid (VOF)
  Method for the Dynamics of Free Boundaries}},\ }\href@noop {} {\bibfield
  {journal} {\bibinfo  {journal} {Journal of Computational Physics}\ }\textbf
  {\bibinfo {volume} {39}},\ \bibinfo {pages} {201} (\bibinfo {year}
  {1981})}\BibitemShut {NoStop}%
\bibitem [{\citenamefont {Jha}\ and\ \citenamefont {Juanes}(2014)}]{Jha2014}%
  \BibitemOpen
  \bibfield  {author} {\bibinfo {author} {\bibfnamefont {B.}~\bibnamefont
  {Jha}}\ and\ \bibinfo {author} {\bibfnamefont {R.}~\bibnamefont {Juanes}},\
  }\bibfield  {title} {\bibinfo {title} {{Coupled multiphase flow and
  poromechanics: A computational model of pore pressure effects on fault slip
  and earthquake triggering}},\ }\href {https://doi.org/10.1002/2013WR015175}
  {\bibfield  {journal} {\bibinfo  {journal} {Water Resources Research}\
  }\textbf {\bibinfo {volume} {50}},\ \bibinfo {pages} {3776} (\bibinfo {year}
  {2014})}\BibitemShut {NoStop}%
\bibitem [{\citenamefont {Kim}\ \emph {et~al.}(2013)\citenamefont {Kim},
  \citenamefont {Tchelepi},\ and\ \citenamefont {Juanes}}]{Kim2013}%
  \BibitemOpen
  \bibfield  {author} {\bibinfo {author} {\bibfnamefont {J.}~\bibnamefont
  {Kim}}, \bibinfo {author} {\bibfnamefont {H.~A.}\ \bibnamefont {Tchelepi}},\
  and\ \bibinfo {author} {\bibfnamefont {R.}~\bibnamefont {Juanes}},\
  }\bibfield  {title} {\bibinfo {title} {{Rigorous coupling of geomechanics and
  multiphase flow with strong capillarity}},\ }\href
  {https://doi.org/10.2118/141268-PA} {\bibfield  {journal} {\bibinfo
  {journal} {SPE Journal}\ }\textbf {\bibinfo {volume} {18}},\ \bibinfo {pages}
  {1123} (\bibinfo {year} {2013})}\BibitemShut {NoStop}%
\bibitem [{\citenamefont {Carrillo}\ \emph {et~al.}(2020)\citenamefont
  {Carrillo}, \citenamefont {Bourg},\ and\ \citenamefont
  {Soulaine}}]{Carrillo2020}%
  \BibitemOpen
  \bibfield  {author} {\bibinfo {author} {\bibfnamefont {F.~J.}\ \bibnamefont
  {Carrillo}}, \bibinfo {author} {\bibfnamefont {I.~C.}\ \bibnamefont
  {Bourg}},\ and\ \bibinfo {author} {\bibfnamefont {C.}~\bibnamefont
  {Soulaine}},\ }\bibfield  {title} {\bibinfo {title} {{Multiphase Flow
  Modeling in Multiscale Porous Media: An Open-Source Micro-Continuum
  Approach}},\ }\href {https://doi.org/10.1016/j.jcpx.2020.100073} {\bibfield
  {journal} {\bibinfo  {journal} {Journal of Computational Physics: X}\
  }\textbf {\bibinfo {volume} {8}},\ \bibinfo {pages} {100073} (\bibinfo {year}
  {2020})}\BibitemShut {NoStop}%
\bibitem [{\citenamefont {Whitaker}(1986)}]{Whitaker1986}%
  \BibitemOpen
  \bibfield  {author} {\bibinfo {author} {\bibfnamefont {S.}~\bibnamefont
  {Whitaker}},\ }\bibfield  {title} {\bibinfo {title} {{Flow in porous media I:
  A theoretical derivation of Darcy's law}},\ }\href
  {https://doi.org/10.1007/BF01036523} {\bibfield  {journal} {\bibinfo
  {journal} {Transport in Porous Media}\ }\textbf {\bibinfo {volume} {1}},\
  \bibinfo {pages} {3} (\bibinfo {year} {1986})}\BibitemShut {NoStop}%
\bibitem [{\citenamefont {Carrillo}\ and\ \citenamefont
  {Bourg}(2020{\natexlab{b}})}]{hybridBiotInterFoam_Code}%
  \BibitemOpen
  \bibfield  {author} {\bibinfo {author} {\bibfnamefont {F.~J.}\ \bibnamefont
  {Carrillo}}\ and\ \bibinfo {author} {\bibfnamefont {I.~C.}\ \bibnamefont
  {Bourg}},\ }\href {https://doi.org/10.5281/ZENODO.4013969} {\bibinfo {title}
  {{hybridBiotInterFoam}}} (\bibinfo {year} {2020}{\natexlab{b}})\BibitemShut
  {NoStop}%
\bibitem [{\citenamefont {Huang}\ \emph {et~al.}(2012)\citenamefont {Huang},
  \citenamefont {Zhang}, \citenamefont {Callahan},\ and\ \citenamefont
  {Ayoub}}]{Huang2012a}%
  \BibitemOpen
  \bibfield  {author} {\bibinfo {author} {\bibfnamefont {H.}~\bibnamefont
  {Huang}}, \bibinfo {author} {\bibfnamefont {F.}~\bibnamefont {Zhang}},
  \bibinfo {author} {\bibfnamefont {P.}~\bibnamefont {Callahan}},\ and\
  \bibinfo {author} {\bibfnamefont {J.~A.}\ \bibnamefont {Ayoub}},\ }\bibfield
  {title} {\bibinfo {title} {{Fluid injection experiments in 2D porous
  media}},\ }\href {https://doi.org/10.2118/140502-PA} {\bibfield  {journal}
  {\bibinfo  {journal} {SPE Journal}\ }\textbf {\bibinfo {volume} {17}},\
  \bibinfo {pages} {903} (\bibinfo {year} {2012})}\BibitemShut {NoStop}%
\bibitem [{\citenamefont {Spearman}(2017)}]{Spearman2017}%
  \BibitemOpen
  \bibfield  {author} {\bibinfo {author} {\bibfnamefont {J.}~\bibnamefont
  {Spearman}},\ }\bibfield  {title} {\bibinfo {title} {{An examination of the
  rheology of flocculated clay suspensions}},\ }\href
  {https://doi.org/10.1007/s10236-017-1041-8} {\bibfield  {journal} {\bibinfo
  {journal} {Ocean Dynamics}\ }\textbf {\bibinfo {volume} {67}},\ \bibinfo
  {pages} {485} (\bibinfo {year} {2017})}\BibitemShut {NoStop}%
\bibitem [{\citenamefont {Quemada}(1977)}]{Quemada1977}%
  \BibitemOpen
  \bibfield  {author} {\bibinfo {author} {\bibfnamefont {D.}~\bibnamefont
  {Quemada}},\ }\bibfield  {title} {\bibinfo {title} {{Rheology of concentrated
  disperse systems and minimum energy dissipation principle}},\ }\href
  {https://doi.org/10.1007/BF01516932} {\bibfield  {journal} {\bibinfo
  {journal} {Rheologica Acta}\ }\textbf {\bibinfo {volume} {16}},\ \bibinfo
  {pages} {82} (\bibinfo {year} {1977})}\BibitemShut {NoStop}%
\bibitem [{\citenamefont {Zhou}\ \emph {et~al.}(2010)\citenamefont {Zhou},
  \citenamefont {Dong}, \citenamefont {{De Pater}},\ and\ \citenamefont
  {Zitha}}]{Zhou2010}%
  \BibitemOpen
  \bibfield  {author} {\bibinfo {author} {\bibfnamefont {J.}~\bibnamefont
  {Zhou}}, \bibinfo {author} {\bibfnamefont {Y.}~\bibnamefont {Dong}}, \bibinfo
  {author} {\bibfnamefont {C.~J.}\ \bibnamefont {{De Pater}}},\ and\ \bibinfo
  {author} {\bibfnamefont {P.~L.}\ \bibnamefont {Zitha}},\ }\bibfield  {title}
  {\bibinfo {title} {{Experimental study of the impact of shear dilation and
  fracture behavior during polymer injection for heavy oil recovery in
  unconsolidated reservoirs}},\ }in\ \href {https://doi.org/10.2118/137656-ms}
  {\emph {\bibinfo {booktitle} {In Proceedings - Canadian Unconventional
  Resources and International Petroleum Conference,Paper Number SPE-137656-MS,
  Society of Petroleum Engineerings}}}\ (\bibinfo {year} {2010})\BibitemShut
  {NoStop}%
\bibitem [{\citenamefont {Leverett}(1941)}]{CLeverett}%
  \BibitemOpen
  \bibfield  {author} {\bibinfo {author} {\bibfnamefont {M.}~\bibnamefont
  {Leverett}},\ }\bibfield  {title} {\bibinfo {title} {{Capillary Behavior in
  Porous Solids}},\ }\href {https://doi.org/10.2118/941152-g} {\bibfield
  {journal} {\bibinfo  {journal} {Transactions of the AIME}\ }\textbf {\bibinfo
  {volume} {142}},\ \bibinfo {pages} {152} (\bibinfo {year}
  {1941})}\BibitemShut {NoStop}%
\bibitem [{\citenamefont {Li}\ and\ \citenamefont {Benson}(2015)}]{Li2015}%
  \BibitemOpen
  \bibfield  {author} {\bibinfo {author} {\bibfnamefont {B.}~\bibnamefont
  {Li}}\ and\ \bibinfo {author} {\bibfnamefont {S.~M.}\ \bibnamefont
  {Benson}},\ }\bibfield  {title} {\bibinfo {title} {{Influence of small-scale
  heterogeneity on upward CO2 plume migration in storage aquifers}},\ }\href
  {https://doi.org/10.1016/j.advwatres.2015.07.010} {\bibfield  {journal}
  {\bibinfo  {journal} {Advances in Water Resources}\ }\textbf {\bibinfo
  {volume} {83}},\ \bibinfo {pages} {389} (\bibinfo {year} {2015})}\BibitemShut
  {NoStop}%
\end{thebibliography}%


%


\begin{thebibliography}{11}
\expandafter\ifx\csname natexlab\endcsname\relax\def\natexlab#1{#1}\fi
\expandafter\ifx\csname url\endcsname\relax
  \def\url#1{\texttt{#1}}\fi
\expandafter\ifx\csname urlprefix\endcsname\relax\def\urlprefix{URL }\fi

\bibitem[{Ahmed et~al.(2007)Ahmed, Karim, Gay, Fanhong, and
  Manoj}]{Abou-Sayed2004}
Ahmed, A.-S., Karim, Z., Gay, W., Fanhong, M., Manoj, S., 2007. {Fracture
  Propagation and Formation Disturbance during Injection and Frac-Pack
  Operations in Soft Compacting Rocks}. In: Proceedings - SPE Annual Technical
  Conference and Exhibition. Society of Petroleum Engineers, pp. 3453--3464.

\bibitem[{Carrillo and Bourg(2019)}]{Carrillo2019a}
Carrillo, F.~J., Bourg, I.~C., 2019. {A Darcy-Brinkman-Biot Approach to
  Modeling the Hydrology and Mechanics of Porous Media Containing Macropores
  and Deformable Microporous Regions}. Water Resources Research 55~(10),
  8096--8121.

\bibitem[{Carrillo and Bourg(2020{\natexlab{a}})}]{hybridBiotInterFoam_Code}
Carrillo, F.~J., Bourg, I.~C., sep 2020{\natexlab{a}}. {hybridBiotInterFoam}.
\newline\urlprefix\url{https://zenodo.org/record/4013969}

\bibitem[{Carrillo and Bourg(2020{\natexlab{b}})}]{Carrillo2020MDBB}
Carrillo, F.~J., Bourg, I.~C., 2020{\natexlab{b}}. Modeling multiphase flow
  within and around deformable porous materials: A darcy-brinkman-biot
  approach. Earth and Space Science Open Archive, 33.
\newline\urlprefix\url{https://www.essoar.org/doi/abs/10.1002/essoar.10504277.2}

\bibitem[{Carrillo et~al.(2020)Carrillo, Bourg, and Soulaine}]{Carrillo2020}
Carrillo, F.~J., Bourg, I.~C., Soulaine, C., sep 2020. {Multiphase Flow
  Modeling in Multiscale Porous Media: An Open-Source Micro-Continuum
  Approach}. Journal of Computational Physics: X 8, 100073.

\bibitem[{Ferrari et~al.(2015)Ferrari, Jimenez-Martinez, {Le Borgne},
  M{\'{e}}heust, and Lunati}]{Ferrari2015}
Ferrari, A., Jimenez-Martinez, J., {Le Borgne}, T., M{\'{e}}heust, Y., Lunati,
  I., 2015. {Challenges in modeling unstable two-phase flow experiments in
  porous micromodels}. Water Resources Research 51~(3), 1381--1400.

\bibitem[{Quemada(1977)}]{Quemada1977}
Quemada, D., jan 1977. {Rheology of concentrated disperse systems and minimum
  energy dissipation principle}. Rheologica Acta 16~(1), 82--94.

\bibitem[{Sandnes et~al.(2011)Sandnes, Flekk{\o}y, Knudsen, M{\aa}l{\o}y, and
  See}]{Sandnes2011a}
Sandnes, B., Flekk{\o}y, E.~G., Knudsen, H.~A., M{\aa}l{\o}y, K.~J., See, H.,
  2011. {Patterns and flow in frictional fluid dynamics}. Nature Communications
  2~(1), 288.

\bibitem[{Spearman(2017)}]{Spearman2017}
Spearman, J., 2017. {An examination of the rheology of flocculated clay
  suspensions}. Ocean Dynamics 67~(3-4), 485--497.

\bibitem[{van Dam et~al.(2002)van Dam, Papanastasiou, and
  de~Pater}]{VanDam2002}
van Dam, D.~B., Papanastasiou, P., de~Pater, C.~J., 2002. {Impact of rock
  plasticity on hydraulic fracture propagation and closure}. SPE Production and
  Facilities 17~(3), 149--159.

\bibitem[{van Genuchten(1980)}]{VanGenutchen1980}
van Genuchten, M.~T., 1980. {A Closed-form Equation for Predicting the
  Hydraulic Conductivity of Unsaturated Soils}. Soil Science Society of America
  Journal 44~(5), 892--898.

\end{thebibliography}

\end{document}


\begin{frontmatter}



\title{Supplementary Materials for: ``Capillary and Viscous Fracturing During Drainage in Porous Media"}

\author[PRINCETONCHEM]{Francisco J. Carrillo}
\author[PRINCETONCIVIL,PEI]{Ian C. Bourg}

\address[PRINCETONCHEM]{Department of Chemical and Biological Engineering, Princeton University, Princeton, NJ, USA}
\address[PRINCETONCIVIL]{Department of Civil and Environmental Engineering, Princeton University, Princeton, NJ, USA}
\address[PEI]{Princeton Environmental Institute, Princeton University, Princeton, NJ, USA}

\begin{abstract}
This Supplementary Material consists of the following:
\begin{enumerate}
 
\item A \hyperref[description]{Section} describing the fully-coupled model for two-phase flow and solid mechanics.

\item A \hyperref[rheologymod]{Section} that outlines the solid's plastic rheology model. 

\item A \hyperref[Numerical Setup]{Section} that delineates the base simulation setup for the cases presented in Figure 2.



\item A \hyperref[PProfile]{Graph} containing normalized fracturing pressure profiles for all fracturing zones. 

\item A \hyperref[fig:nonUniformPc]{Figure} that depicts the effects of non-uniform capillary entry pressures on invasion-fracturing patterns. 

\item A \hyperref[NPorosity]{Graph} showing the fracturing numbers' dependence on porosity. 

\item A \hyperref[nom]{Table} defining all the symbols used throughout the text. 

\end{enumerate}

\end{abstract}

\end{frontmatter}


\section{Complete Description of Modelling Framework} \label{description}

The complete set of equations in the multiphase Darcy-Brinkman-Biot framework now follows. The combination of these solid and fluid conservation equations leads to a model that tends towards multiphase Navier-Stokes in solid-free regions and towards Biot Theory in porous regions, as described in Figure 1 in the main text.
\begin{equation}\label{f_mass}
\begin{aligned}
\frac{\partial {\phi }_f}{\partial t}+\nabla\cdot\boldsymbol{U}_f=0
\end{aligned}
\end{equation}

\begin{equation}\label{f_sat}
\begin{aligned}
\frac{\partial {\phi }_f{\alpha }_w}{\partial t}+\nabla \cdot\left({\alpha }_w{\boldsymbol{U}}_f\right)+\nabla \cdot\left({\phi }_f{\alpha }_w{\alpha }_n{\boldsymbol{U}}_r\right)=0
\end{aligned}
\end{equation}

\begin{equation}\label{f_mom}
\begin{split}
\frac{\partial {\rho }_f{\boldsymbol{U}}_f}{\partial t}+\nabla \cdot\left(\frac{{\rho }_f}{{\phi }_f}{\boldsymbol{U}}_f{\boldsymbol{U}}_f\right)=&-{\phi }_f\nabla p+{\phi }_f{\rho }_f\boldsymbol{g}+\nabla\cdot\boldsymbol{S} \\ &-{\phi }_f\mu k^{-1}\left({\boldsymbol{U}}_f-{\boldsymbol{U}}_s\right)+{\phi }_f{\boldsymbol{F}}_{c,1}+{\phi }_f{\boldsymbol{F}}_{c,2}\ 
\end{split}
\end{equation}

\begin{equation}\label{s_mass}
\begin{aligned}
\frac{\partial {\phi }_s}{\partial t}+\nabla \cdot\left({\phi }_s{\boldsymbol{U}}_s\right)=0
\end{aligned}
\end{equation}

\begin{equation}\label{s_mom}
\begin{aligned}
-\nabla\cdot\overline{\boldsymbol{\sigma }}=-{\phi }_s\nabla p +{{\phi }_s\rho }_s\boldsymbol{g}+{\phi }_f\mu k^{-1}\left({\boldsymbol{U}}_f-{\boldsymbol{U}}_s\right)-{\phi }_f{\boldsymbol{F}}_{c,1}+{\phi }_s{\boldsymbol{F}}_{c,2}
\end{aligned}
\end{equation}

We now state the closed-form expressions of the multiscale parameters $\mu k^{-1}\mathrm{, }\ {\boldsymbol{F}}_{c,i}\mathrm{, and}\ {\boldsymbol{U}}_{r}$, which are defined differently in each region. A full derivation and discussion of these parameters can be found in \cite{Carrillo2020} and \cite{Carrillo2020MDBB}.

\begin{equation}\label{uk_def}
\mu k^{-1}=
    \begin{cases}
        0 & \textnormal{in solid-free regions}\\
        k^{-1}_0{\left(\frac{k_{r,w}}{{\mu }_w}+\frac{k_{r,n}}{{\mu }_n}\right)}^{-1} &  \textnormal{in porous regions}
    \end{cases}
\end{equation}

\begin{equation}\label{Fc1_def}
{\boldsymbol{F}}_{c,1}=
    \begin{cases}
        -\frac{\gamma}{{\phi}_f}\nabla\cdot \left({\boldsymbol{n}}_{w,n}\right)\nabla{\alpha}_{w} & \textnormal{in solid-free regions}\\
        -p_c\nabla{\alpha }_w &  \textnormal{in porous regions}
    \end{cases}
\end{equation}

\begin{equation}\label{Fc2_def}
{\boldsymbol{F}}_{c,2}=
    \begin{cases}
        0 & \textnormal{in solid-free regions}\\
        M^{-1}\left(M_w{\alpha}_n-M_n{\alpha}_w\right)\left(\nabla p_c+\left({\rho}_w-{\rho}_n\right)\boldsymbol{g}\right) &  \textnormal{in porous regions}
    \end{cases}
\end{equation}

\begin{equation}\label{nw_def}
{\boldsymbol{n}}_{w,n}=
    \begin{cases}
        \frac{\nabla{\alpha}_{w}}{\left|\nabla{\alpha }_{w}\right|} & \textnormal{in solid-free regions}\\
        {cos \left(\theta \right)\ }{\boldsymbol{n}}_{wall}+{sin \left(\theta \right)\ }{\boldsymbol{t}}_{wall} &  \textnormal{at the interface between solid-free porous regions}
    \end{cases}
\end{equation}

\begin{equation}\label{Ur_def}
{\boldsymbol{U}}_{r}=
    \begin{cases}
        C_{\alpha \ }max\left(\left|{\boldsymbol{U}}_f\right|\right)\frac{\nabla{\alpha }_w}{\left|\nabla{\alpha }_w\right|} & \textnormal{in solid-free regions}\\
        {\phi }^{-1}\left[ \begin{array}{c}
-\left(M_w{\alpha }^{-1}_w-M_n{\alpha }^{-1}_n\right)\nabla p + \left({\rho }_wM_w{\alpha }^{-1}_w-{\rho }_wM_n{\alpha }^{-1}_n\right)\boldsymbol{g}+ \\ 
\left(M_w{\alpha }_n{\alpha }^{-1}_w +M_n{\alpha }_w{\alpha }^{-1}_n\right)\nabla p_c- \left(M_w{\alpha }^{-1}_w- M_n{\alpha }^{-1}_n\right)p_{c}\nabla{\alpha }_w \end{array}\right] &  \textnormal{in porous regions}
\end{cases}
\end{equation}

\noindent where $C_{\alpha }$ is an interface compression parameter (traditionally set to values between 1 and 4 in the Volume-of-Fluid method), $k_0$ is the absolute permeability, $k_{r,i}$ and $M_i=k_0k_{i,r}/{\mu }_i\ $are the relative permeability and mobility of each fluid, and $M=M_w+M_n$. Lastly, $\theta $ is the imposed contact angle at the porous wall, and $\boldsymbol{n}_{wall}$  and $\boldsymbol{t}_{wall}$ are the normal and tangential directions relative to said wall, respectively.

Closure of this system of equations requires appropriate constitutive models describing the averaged behavior of the different phases within the porous regions. In the present paper we use the following well established constitutive models: absolute permeability is modeled as isotropic and porosity-dependent through the Kozeny-Carman relation ($k_0 = k_0^0 \frac{\phi_f^3}{(1-\phi_f)^2}$); relative permeabilities and average capillary pressures within the porous domains are represented using the well-known Van Genutchen \citep{VanGenutchen1980} model; plasticity is described through the Herschel-Bulkley model, were the solid viscously deforms only after local stresses become higher than the material yield stress; finally the solid's yield stress and plastic viscosity are modeled as solid fraction-dependent based on the Quemada fractal model \citep{Quemada1977,Spearman2017}. For the reader's convenience, a full implementation of this framework and its related models are included in the accompanying code \citep{hybridBiotInterFoam_Code}.

\newpage

\section{Solid Rheology Models}\label{rheologymod}

\subsection{Hershel-Bulkley Plasticity} \label{rheology_Herschel}

A Bingham plastic is a material that deforms only once it is under a sufficiently high stress. After this yield stress is reached, it will deform viscously and irreversibly. The Herschel-Bulkley rheological model combines the properties of a Bingham plastic with a power-law viscosity model, such that said plastic can be shear thinning or shear thickening during deformation. In OpenFOAM{\circledR} this model is implemented as follows: 

\[\boldsymbol{\sigma }={\mu }^{eff}_s\left(\ \nabla {\boldsymbol{U}}_s+{ \ \left(\nabla {\boldsymbol{U}}_s\right)}^T-\frac{2}{3}\nabla \cdot\left({\boldsymbol{U}}_s\mathrm{\ }\boldsymbol{I}\right)\right)\mathrm{\ }\] 

\noindent where ${\mu }^{eff}_s$ is the effective solid plastic viscosity, which is then modeled through a power law expression: 

\[{\mu }^{eff}_s={\mathrm{min} \left({\mu }^0_s\mathrm{\ ,\ \ \ }\frac{\tau }{\eta }+{\mu }_s{\eta}^{n-1}\right)\ }\]      

\noindent where$\ {\mu }^0_s$ is the limiting viscosity (set to a large value), $\tau \ $is the yield stress, ${\mu }_s$ is the viscosity of the solid once the yield stress is overcome, $n$ is the flow index ($n=1$ for constant viscosity), and $\eta $ is the shear rate.       

\subsection{Quemada Rheology Model} \label{rheology_Quemada}

The Quemada rheology model \cite{Quemada1977,Spearman2017} is a simple model that accounts for the fact that the average yield stress and effective viscosity of a plastic are functions of the solid fraction. These two quantities are large at high solid fractions and small at low solid fractions, as described by the following relations

\[\tau ={\tau }_0{\left(\frac{\left({{\phi }_s}/{{\phi }^{max}_s}\right)}{\left(1-{{\phi }_s}/{{\phi }^{max}_s}\boldsymbol{\ }\right)}\right)}^D\] 

\[{\mu }_s=\frac{{\mu }_0}{{\left(1-\frac{{\phi }_s}{{\phi }^{max}_s}\right)}^2}\] 

\noindent here, ${\phi }^{max}_s$ is the maximum solid fraction possible (perfect incompressible packing),$\ {\tau }_0$ is the yield stress at ${\phi }_s={\phi }^{max}_s/2$ ,$\ {\mu }_0$ is the viscosity of the fluid where the solid would be suspended at low solid fractions (high fluid fractions), and $D$ is a scaling parameter based on the solid's fractal dimension. 

\newpage

\section{General Methodology for Fracturing Simulations shown in Figure 2} \label{Numerical Setup}

As shown in our previous work \citep{Carrillo2020MDBB}, the simulations shown in Figure 2 where implemented as follows: Numerical parameters were set to the known properties of aqueous
glycerin, air, and sand (${\rho }_{gly}=1250\ \mathrm{kg/m^3}$, ${\mu }_{gly}=5\ $to $176\
\mathrm{cP}$, ${\rho }_{air}=1\ \mathrm{kg/m^{3}}$, $\mu
_{air}=0.017\ \mathrm{cP}$, ${\rho}_s=2650\ \mathrm{kg/m^3}$), the air-glycerin surface tension $\gamma =0.063\
\mathrm{kg/s^{2}}$, and the sand grain radius $r_s=100
\ \mu \mathrm{m}$. To mimic the existence of sub-REV scale heterogeneity, the solid fraction was initialized as a normally-distributed field ${\phi}_s=0.64 \pm
0.05$. To account for the non-reversible and compressive nature of the experiments,
the deformable solid was modeled as a Hershel-Bulkley-Quemada plastic with a density-normalized yield stress of $\tau
_{yield}=16.02 \ \mathrm{m^{2}/s^{2}}$ \cite{Abou-Sayed2004,VanDam2002,Carrillo2020MDBB}. Permeability was
modeled as a function of porosity through the Kozeny-Carman relation:
$k=k_0{\phi}^3_f{\phi}^{-2}_s$ with $k_0=6.7\times{10}^{-12}\ \mathrm{m^2}.$ These two complementary models couples the solid rheology and absolute permeability to its porosity, making it harder to deform whenever its compressed and easier to flow through whenever it expands (and vice-versa). Relative permeabilities were calculated through the Van Genuchten model \cite{VanGenutchen1980} with wettability parameter $m = 0.99$; capillary effects were assumed negligible. 

The resulting simulations were implemented in OpenFOAM{\circledR} within a 500 by 500 grid with
constant flow boundary conditions at the inlet, zero-gradient flow conditions at the
outlet, and no-slip boundary conditions for solid displacement at all boundaries. To allow
for a proper comparison between 2-D simulations and 3-D experiments in a Hele-Shaw cell of
thickness $a$, an additional drag term ($12\mu
a^{-2}{\boldsymbol{U}}_f$) was included in the fluid momentum equation \cite{Ferrari2015}. Lastly, the boundary effects of sliding friction and vertical confinement on the solid were neglected, a reasonable simplification that becomes significant at relatively low initial packing fractions ($\phi_s \sim 0.35$) or high confining pressures \cite{Sandnes2011a}. If necessary, these effects could potentially be included into the model by including an additional drag term and a confining pressure into Eqn. 5 in the main text, as done in \cite{Carrillo2019a}

\newpage

\newpage

\section{Fracturing Pressure Profiles} \label{PProfile}

\begin{figure}[htb!]
\begin{center}
\includegraphics[width=0.97\textwidth]{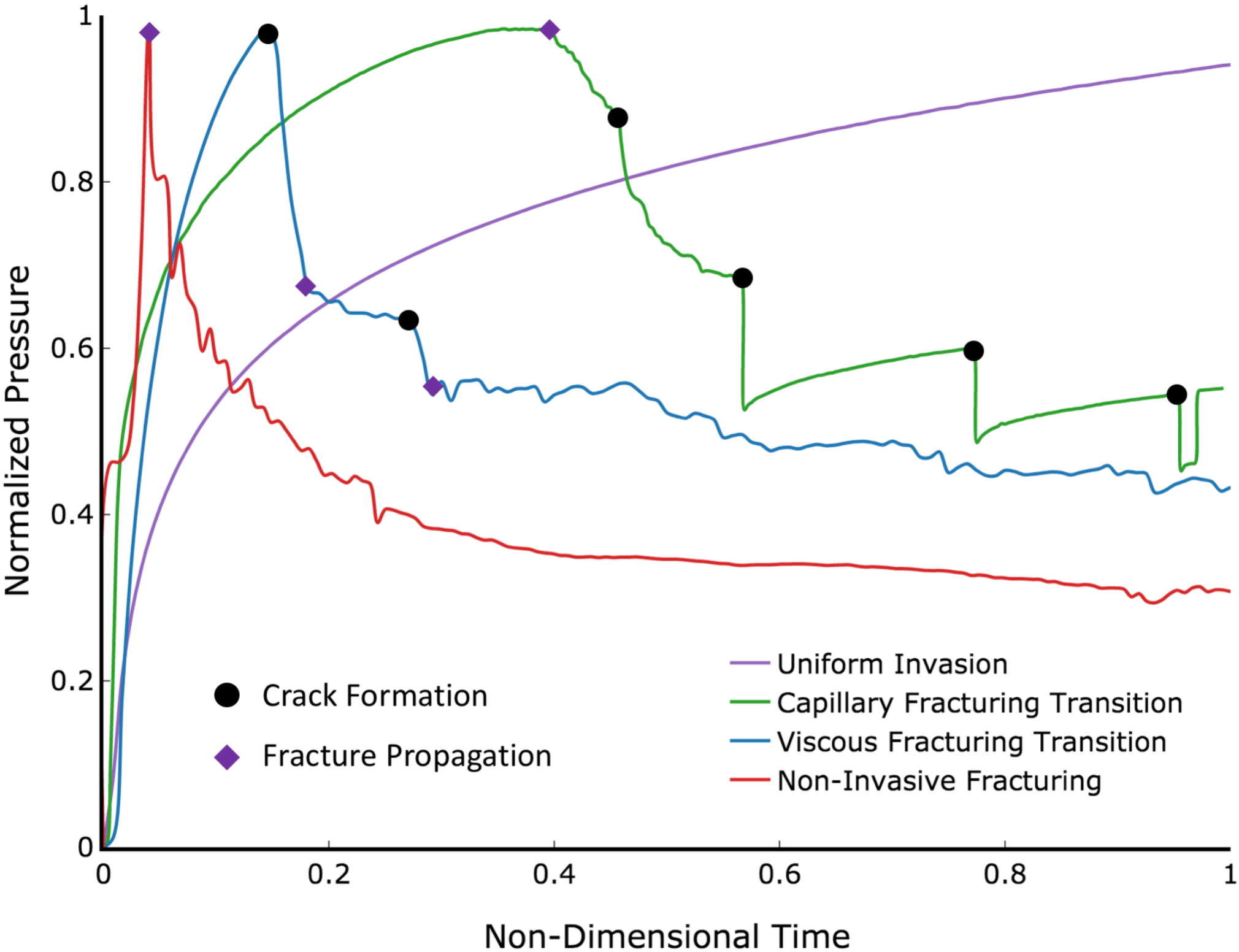}
\caption{Normalized pressure profiles far all four fracturing zones. Note how the two fracturing transition zones exhibit a combination of the characteristic behaviour shown in the Uniform Invasion and Non-Invasive Fracturing pressure curves. Symbols indicate the initiation of non-flow-bearing cracks and flow-bearing fractures. All pressure profiles where normalized by the maximum pressure achieved by each simulation and by the time it took for the fluid-fluid interface to reach the outer boundary in each simulation. \label{PressureP}} 
\end{center} 
\end{figure}

\newpage

\section{Effects of a Non-Uniform Capillary Entry Pressure on the Different Fracturing-Invasion Regimes}\label{fig:nonUniformPc}

\begin{figure}[htb]
\begin{center}
\includegraphics[width=0.97\textwidth]{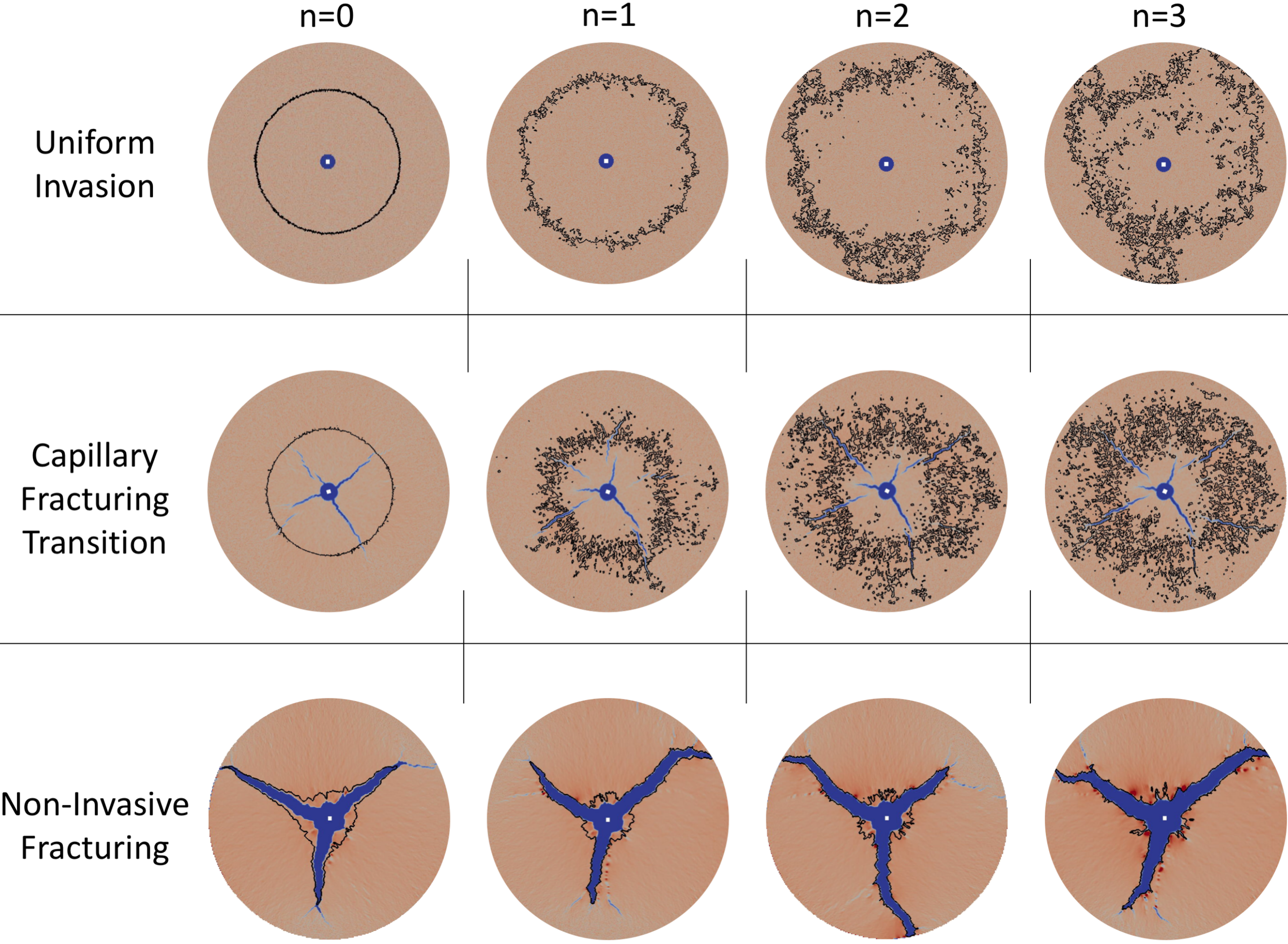}
\caption{Effects of a non-uniform capillary entry pressure distribution on the different fracturing-invasion regimes. Local capillary pressures are described by $p_{c,0} = p_{c,0}^*(\phi_s/\phi_s^{avg})^n$, where $p_{c,0}^*$ is the capillary pressure at $\phi_s=\phi_s^{avg}$, and $n>0$ is a sensitivity parameter. Note how increasing the values of $n$ stimulates the creation of finger-like instabilities at the fluid invasion front and leads to more pronounced nucleation of small fractures at said position. The simulations' color scheme is the same as in Fig. 2 in the main text. }
\end{center} 
\end{figure}

\newpage

\section{Fracturing Numbers Dependence on Porosity} \label{NPorosity}

\begin{figure}[htb!]
\begin{center}
\includegraphics[width=0.7\textwidth]{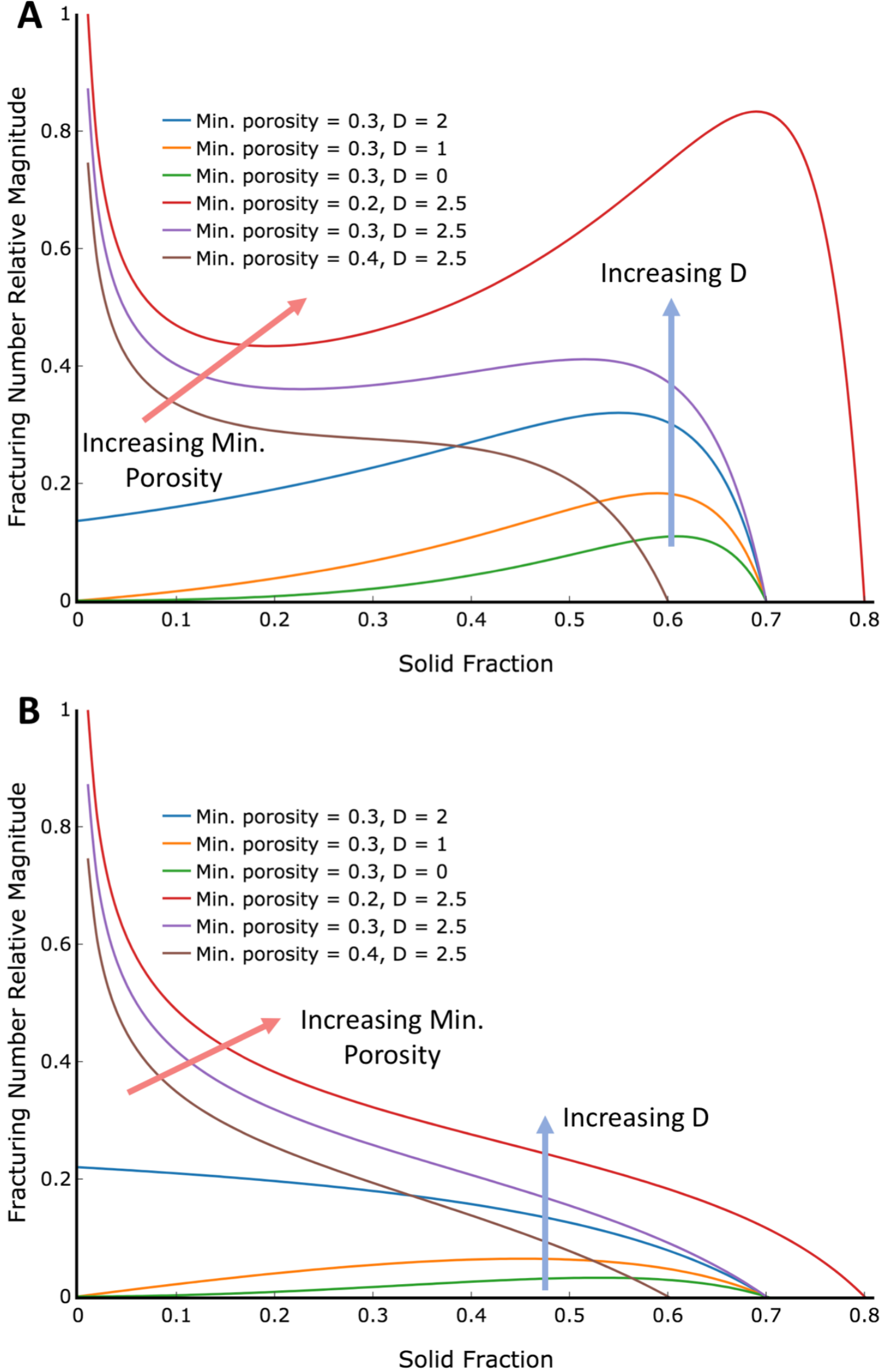}
\caption{Fracturing Numbers' Dependence on Porosity. A) For $N_{vF}$ B) For $N_{cF}$. Colored arrows represent the overall trends in the fracturing number behaviours when changing the fractal parameter ``D" and the minimum porosity (highest possible compression state) parameter in the rheology model. Note the appearance of ``valleys" at high values of D in sub-figure A.  \label{N_Dependence_graph}} 
\end{center} 
\end{figure}

\newpage

\nomenclature{$\rho_i$}{Density of phase $i$ ($\unit{kg/m^3}$) }%
\nomenclature{$\rho_f$}{Single-field fluid density ($\unit{kg/m^3}$) }%
\nomenclature{$\boldsymbol{U}_f$}{Single-field fluid velocity in the grid-based domain ($\unit{m/s}$)}%
\nomenclature{$\boldsymbol{U}_r$}{Relative fluid velocity in the grid-based domain ($\unit{m/s}$)}%
\nomenclature{$\boldsymbol{U}_s$}{Solid velocity in the grid-based domain ($\unit{m/s}$)}%
\nomenclature{$\boldsymbol{v}_{i,j}$}{Velocity of the i-j interface in the continuous physical space ($\unit{m/s}$)}%
\nomenclature{$p$}{Single-field fluid pressure in the grid-based domain ($\unit{Pa}$)}%
\nomenclature{$p_c$}{Capillary pressure ($\unit{Pa}$) }%
\nomenclature{$\boldsymbol{S}$}{Single-field fluid viscous stress tensor in the grid-based domain ($\unit{Pa}$)}%
\nomenclature{$\boldsymbol{\sigma}$}{Elastic (or plastic) solid stress tensor in the grid-based domain ($\unit{Pa}$)}%
\nomenclature{$\tau_0$}{Plastic yield stress ($\unit{Pa}$)}%
\nomenclature{$\gamma$}{Fluid-fluid interfacial tension ($\unit{Pa.m}$)}%
\nomenclature{$\phi_f$}{Porosity field}%
\nomenclature{$\phi_s$}{Solid fraction field}%
\nomenclature{$\alpha_w$}{Saturation of the wetting phase}%
\nomenclature{$\alpha_n$}{Saturation of the non-wetting phase}%
\nomenclature{$\mu_i$}{Viscosity of phase $i$ ($\unit{Pa.s}$) }%
\nomenclature{$\mu_f$}{Single-field viscosity ($\unit{Pa.s}$) }%
\nomenclature{$D$}{Solid Fractal Rheological Parameter}%
\nomenclature{$k$}{Apparent permeability ($\unit{m^2}$) }%
\nomenclature{$\boldsymbol{g}$}{Gravity vector ($\unit{m.s^{-2}}$) }%
\nomenclature{$\boldsymbol{F}_{c,i}$}{Surface tension force in the grid-based domain ($\unit{Pa.m^{-1}}) $}%
\nomenclature{$C_{\alpha}$}{Parameter for the compression velocity model}%
\nomenclature{$k_0$}{Absolute permeability ($\unit{m^2}$) }%
\nomenclature{$k_{r,i}$}{Relative permeability with respect to phase $i$ }%
\nomenclature{$M_i$}{Mobility of phase $i$ ($\unit{kg^{-1}m^3s^{-1}}$)}%
\nomenclature{$M$}{Total mobility ($\unit{kg^{-1}m^3s^{-1}}$)}%
\nomenclature{$\theta$}{Surface contact angle}%
\nomenclature{$\boldsymbol{n}_{wall}$}{Normal vector to the porous surface}%
\nomenclature{$\boldsymbol{t}_{wall}$}{Tangent vector to the porous surface}%
\nomenclature{$p_{c,0}$}{Entry capillary pressure ($\unit{Pa}$)}%
\nomenclature{$\boldsymbol{\tau}$}{Terzaghi stress tensor in the grid-based domain ($\unit{Pa}$)}%

\printnomenclature\label{nom}

\section*{References}
\bibliographystyle{elsarticle-harv}
\bibliography{ref,Sibib}